\documentclass[12pt]{article}
\usepackage{mathbbol}
\def\beq{\begin{equation}}
\def\eeq{\end{equation}}
\def\beqa{\begin{eqnarray}}
\def\eeqa{\end{eqnarray}}
\def\Pexp{{\rm Pexp}}
\def\adj{{\rm adj\,}}
\def\sign{{\rm sign}}
\def\cO{{\cal{O}}}
\def\cF{{\cal{F}}}
\def\cV{{\cal{V}}}
\def\cR{{\cal{R}}}
\def\cS{{\cal{S}}}
\def\cZ{{\cal{Z}}}
\def\k{{\bf k}}
\def\U{{\hat U}}
\def\Y{{\hat Y}}
\def\sn{{\rm sn}}
\def\cn{{\rm cn}}
\def\dn{{\rm dn}}
\def\ff{{\rm{ff}}}
\def\zm{{\rm{zm}}}
\def\mon{{\rm{mon}}}
\def\ddz{{\frac{d}{dz}}}
\def\Tr{{\rm Tr}}
\def\tr{{\rm tr}}
\def\diag{{\rm diag}}
\def\pl{{{\cal P}_\infty}}
\newcommand{\refeq}[1]{\mbox{Eq.~(\ref{eq:#1})}}
\newcommand{\half}{{\scriptstyle{{1\over 2}}}}
\newcommand{\third}{{\scriptstyle{{1\over 3}}}}
\newcommand{\quart}{{\scriptstyle{{1\over 4}}}}
\newcommand{\veps}{\varepsilon}
\newcommand{\vphi}{\varphi}
\newcommand{\ein}{{\Eins}}

\begin{document}
\begin{center}{\Large\bf Higher charge calorons with non-trivial 
holonomy}\\[1cm]
{\bf Falk Bruckmann, D\'aniel N\'ogr\'adi} and {\bf Pierre van Baal}\\[3mm]
{\em Instituut-Lorentz for Theoretical Physics, University
of Leiden,\\ P.O.Box 9506, NL-2300 RA Leiden, The Netherlands}
\end{center}

\section*{Abstract}
The full ADHM-Nahm formalism is employed to find exact higher charge
caloron solutions with non-trivial holonomy, extended beyond the axially 
symmetric solutions found earlier. Particularly interesting is the case 
where the constituent monopoles, that make up these solutions, are not 
necessarily well-separated. This is worked out in detail for charge 2. 
We resolve the structure of the extended core, which was previously 
localized only through the singularity structure of the zero-mode density 
in the far field limit. We also show that this singularity structure 
agrees exactly with the abelian charge distribution as seen through 
the abelian component of the gauge field. As a by-product zero-mode 
densities for charge 2 magnetic monopoles are found.

\section{Introduction}\label{sec:intro}

Calorons are instantons at finite temperature. For a long time the influence of
a background Polyakov loop on the properties of these topological excitations 
has been neglected. Solutions were constructed long ago~\cite{HaSh} and were 
studied in detail in the semi-classical approximation~\cite{GPY}. In all these 
studies the Polyakov loop at spatial infinity (also called the holonomy) was 
trivial, i.e. an element of the center of the gauge group. That the influence 
of the background Polyakov loop on the topological excitations can be dramatic 
is particularly clear in the confined phase, where on average its trace 
vanishes. Caloron solutions in such backgrounds were constructed only 
relatively recently~\cite{KvB,Lee} and can be seen as composed of massive 
monopole constituents with their magnetic charges adding to zero. 

It was observed that the one-loop correction to the action for configurations 
with a non-trivial asymptotic value of the Polyakov loop gives rise to an 
infinite action barrier, which were therefore considered irrelevant~\cite{GPY}. 
However, the infinity simply arises due to the integration over the finite 
energy {\em density} induced by the perturbative fluctuations in the background
of a non-trivial Polyakov loop~\cite{Weiss}. The proper setting would therefore
rather be to calculate the non-perturbative contribution of calorons (with a 
given asymptotic value of the Polyakov loop) to this energy density, as was 
first successfully implemented in supersymmetric theories~\cite{Khoze}, where 
the perturbative contribution vanishes. It has a minimum where the trace of 
the Polyakov loop vanishes, i.e. at maximal non-trivial holonomy.

In a recent study at high temperatures, where one presumably can trust the 
semi-classical approximation, the non-perturbative contribution of these 
monopole constituents (also called dyons) was computed~\cite{Diak}. When 
added to the perturbative contribution~\cite{Weiss} with its minima at center 
elements, a local minimum develops where the trace of the Polyakov loop 
vanishes, deepening further for decreasing temperature. This gives support 
for a phase in which the center symmetry, broken in the high temperature 
phase, is restored and provides an indication that the monopole constituents 
are the relevant degrees of freedom for the confined phase.

Also lattice studies, both using cooling~\cite{Berlin} and chiral 
fermion zero-modes~\cite{Gatt} as filters, have now confirmed that monopole 
constituents do dynamically occur in the confined phase. A charge 1 caloron 
is seen for SU($n$) to consist of $n$ constituent monopoles. In the deconfined 
phase, due to the fact that the average Polyakov loop becomes a center element, 
the caloron returns to the form known as the Harrington-Shepard 
solution~\cite{HaSh}. The latter can also be interpreted as consisting 
of constituent monopoles, however, with $n-1$ of them being massless.

To be precise, for self-dual configurations in the background of non-trivial 
holonomy the masses of constituent monopoles are given by $8\pi^2\nu_j/\beta$, 
with $\nu_j\equiv\mu_{j+1}-\mu_j$. The $\mu_i$ are related to the eigenvalues 
of the Polyakov loop at spatial infinity, 
\beq
\pl=\lim_{|\vec x|\to\infty}\Pexp(\int_0^\beta A_0(t,\vec x)dt)=g^\dagger
\exp(2\pi i\diag(\mu_1,\mu_2,\ldots,\mu_n))g,
\eeq
(this expression assumes the periodic gauge $A_\mu(t,\vec x)=A_\mu(t+\beta,
\vec x)$) where $g$ is the gauge rotation used to diagonalize $\pl$ and 
$\beta$ is the period in the imaginary time direction, related to the 
inverse temperature. The eigenvalues $\exp(2\pi i\mu_j)$ are to be ordered 
on the circle such that $\mu_1\leq\mu_2\leq\ldots\leq\mu_n\leq\mu_{n+1}$, 
with $\mu_{n+j}\equiv 1+\mu_j$ and $\sum_{i=1}^n\mu_i=0$, which guarantees 
that the masses add up to $8\pi^2/\beta$, the instanton action per unit 
(imaginary) time. At higher topological charge $k$, the parameter space
of widely separated constituent monopoles is described by $kn$ monopole
constituents, $k$ of each of the $n$ types of abelian charges (with overall
charge neutrality).

We established in an earlier paper~\cite{Us} that well-separated constituents 
act as point sources for the so-called far field (that is far removed from any 
of the cores). When constituents of opposite charge ($n$ constituents of 
different type) come together, the action density no longer deviates 
significantly from that of a standard instanton. Its scale parameter $\rho$ 
is related to the constituent separation $d$ through $\rho^2/\beta\approx d$. 
Yet, the gauge field is vastly different, as is seen from the fact that 
within the confines of the peak there are $n$ locations where two of the 
eigenvalues of the Polyakov loop coincide~\cite{Poly,PolyL}, thus in some 
sense varying over the maximal range available (e.g. for SU(2) from $\Eins_2$ 
to $-\Eins_2$), whereas for trivial holonomy only one such location is found.

\begin{figure}[htb]
\vspace{9.3cm} 
\includegraphics{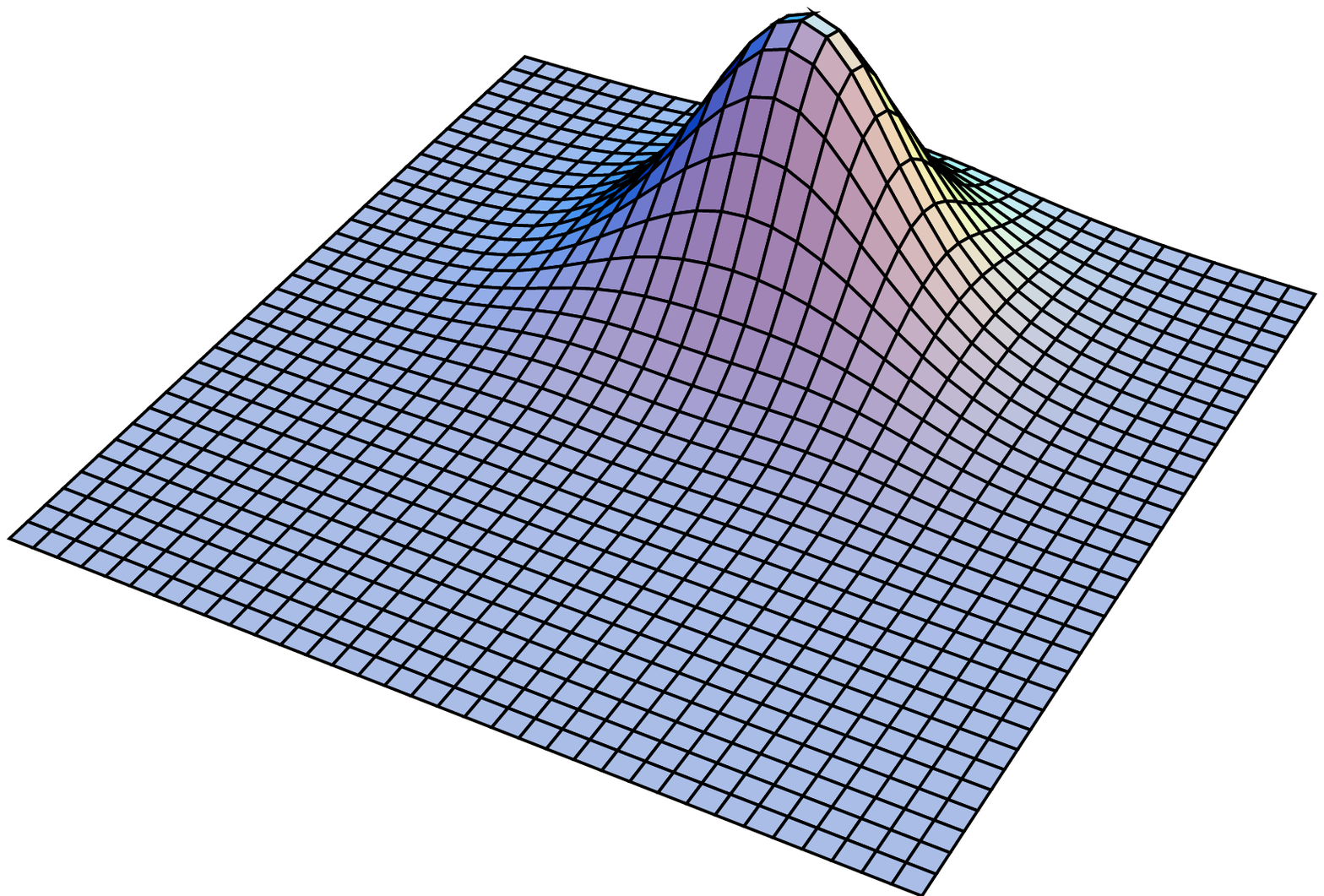}
\includegraphics{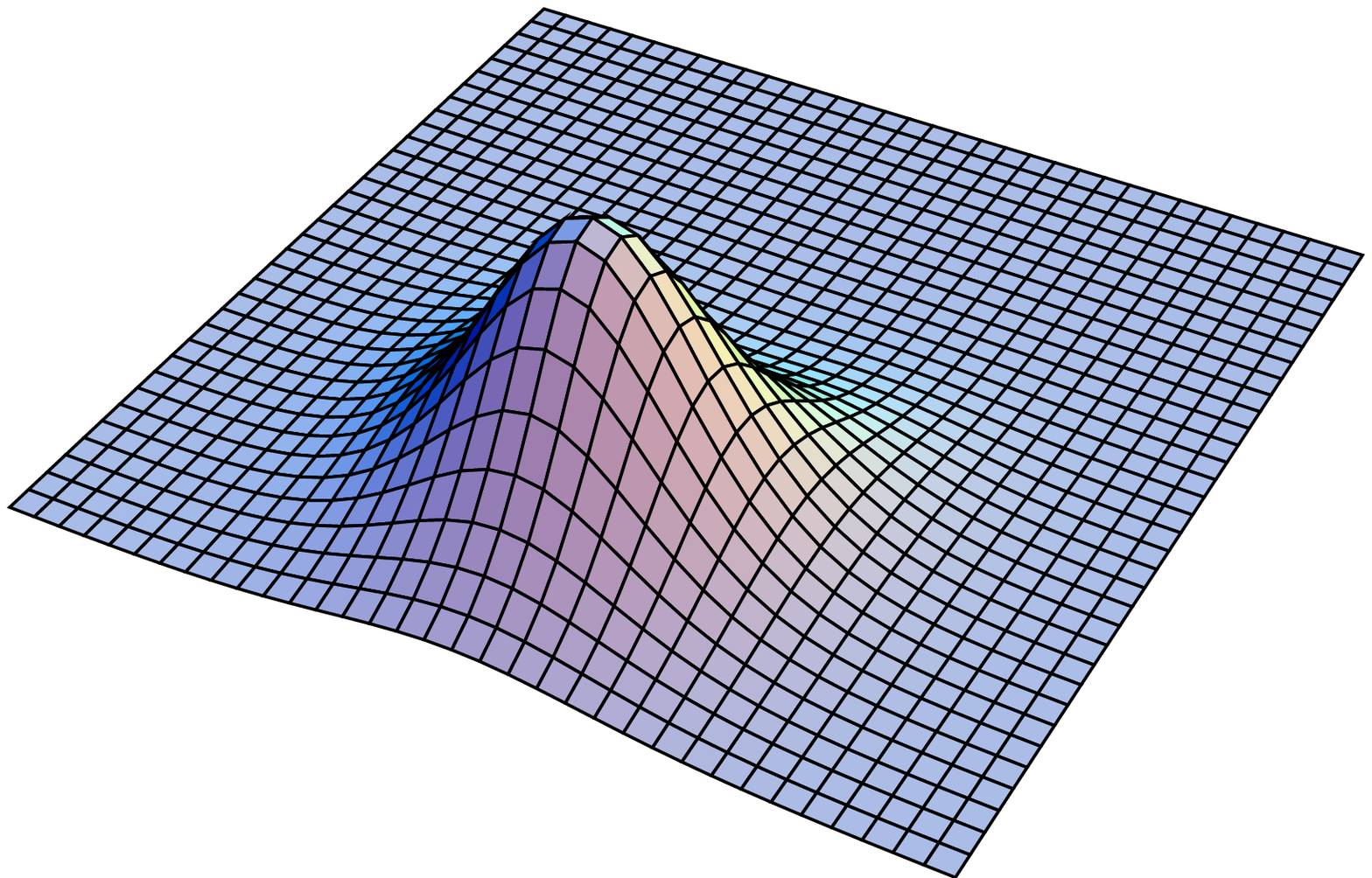}
\includegraphics{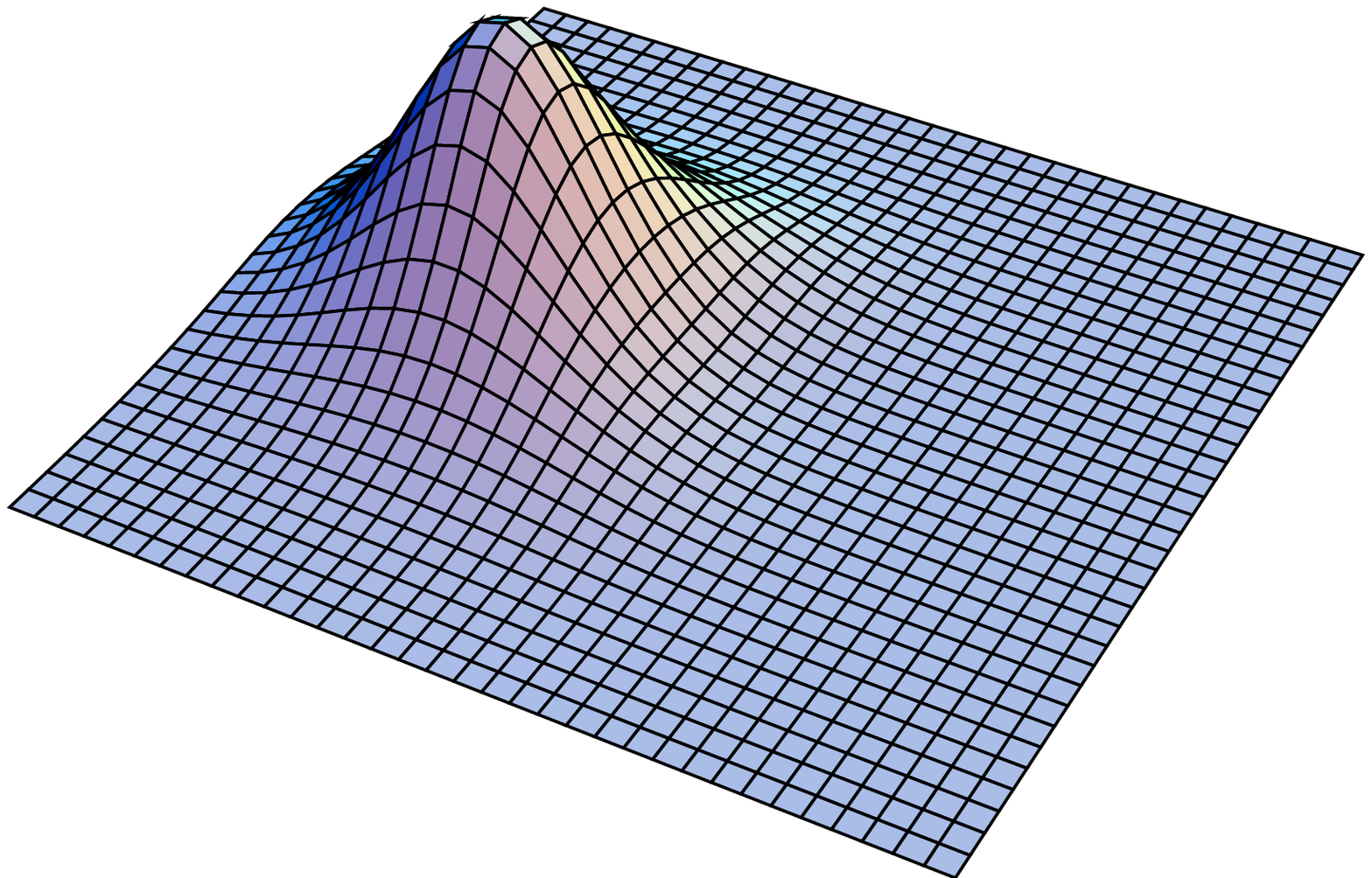}
\includegraphics{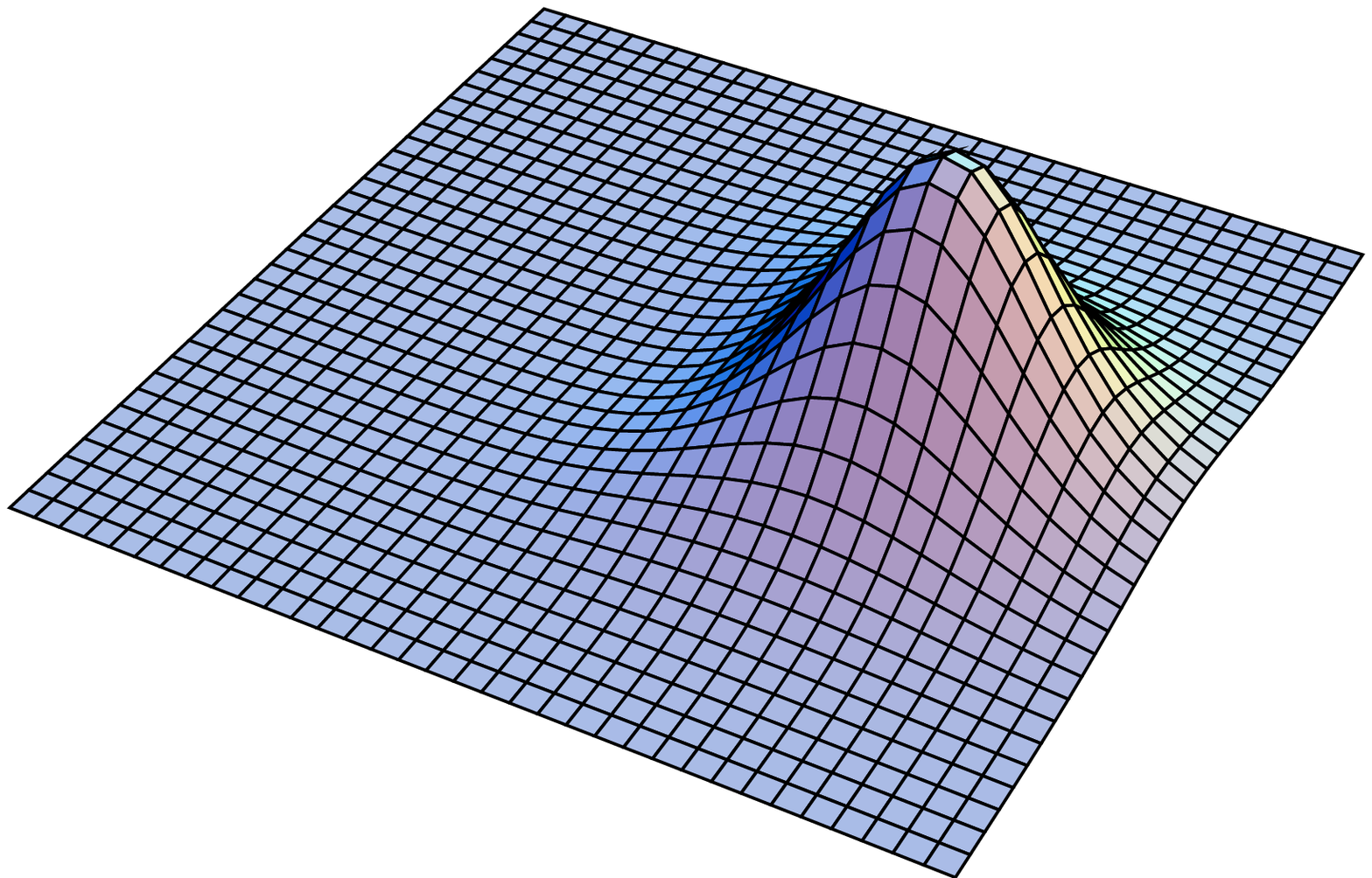}
\includegraphics{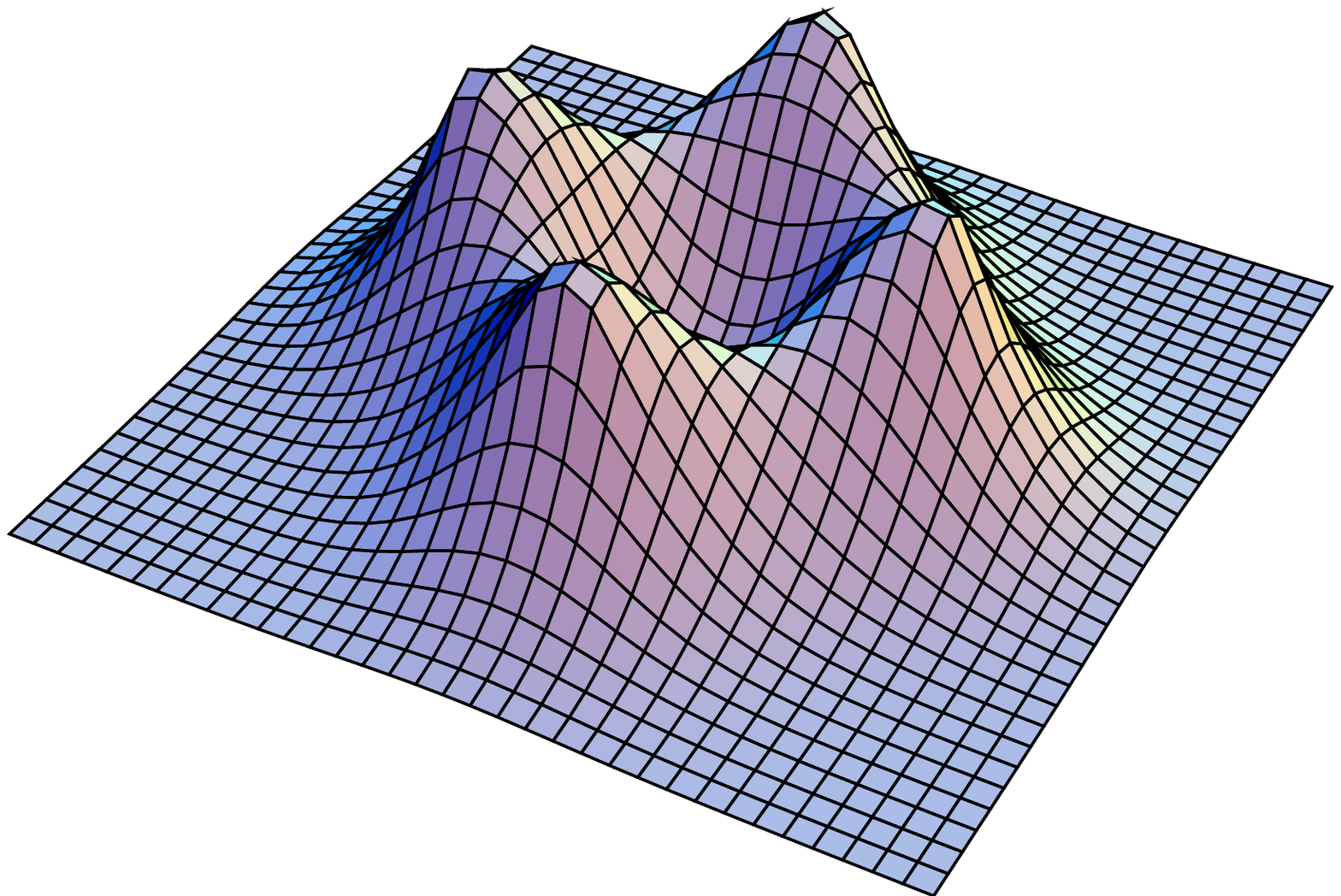}
\caption{In the middle is shown the action density in the plane of the
constituents at $t=0$ for an SU(2) charge 2 caloron with $\tr\,\pl=0$, in 
the regime where constituents are not well-separated. On a scale enhanced 
by a factor $10\pi^2$ are shown the densitities for the two zero-modes, using 
either periodic (left) or anti-periodic (right) boundary conditions in the 
time direction. This solution is for the so-called ``crossed" configuration 
with $\k=0.997$ and $D=8.753$, see Sect.~\ref{sec:plots} for more 
details.}\label{fig:fig1} 
\end{figure}

On the other hand, when constituents of equal charge come together typically 
an extended core structure is found. This was deduced, in particular for the 
case of charge 2 calorons, from our ability to analytically determine the 
zero-mode density (summed over the two zero-modes implied by the index theorem)
in the far field limit, neglecting exponential contributions\footnote{This is 
in some sense equivalent to the high temperature limit, with constituent 
masses given by $8\pi^2\nu_m/\beta$, such that the range of the exponential 
contributions shrinks inversely proportional with the temperature.}. In this 
limit it forms a singular distribution on a disc bounded by an ellipse, but 
approaches two delta functions for well-separated like-charge constituents. 
This zero-mode density only sees constituents of a given charge, depending on 
the boundary condition for the fermions in the imaginary time direction, which 
can be chosen to be a U(1) phase (containing the physically relevant choice of 
anti-periodic boundary conditions for thermal field theory). We will show for 
SU(2) that their difference for periodic and anti-periodic boundary conditions 
coincides {\em exactly} with the (abelian) charge distribution extracted 
from the gauge field in the same limit, making contact with an old result 
due to Hurtubise~\cite{Hurt} for the asymptotic behaviour of the monopole 
Higgs field. 

We found two particular parametrizations within the SU(2) charge 2 moduli 
space that exhibited these extended charge distributions. The first includes
as a limit arbitrary charge 2 monopoles. The second of these parametrizations 
contains as a limiting case the axially symmetric configurations constructed 
for arbitrary charge in Ref.~\cite{BrvB}. Deforming away from the axial 
configuration the two discs overlap. Describing the intricate behaviour for 
the non-abelian core of these configurations in this region of the parameter 
space requires one to find exact solutions, which are presented here. We rely 
on early work of Nahm~\cite{NCal} and Panagopoulos~\cite{Hari} for charge 2 
magnetic monopoles, which is simplified to some extent by our formalism that 
uses the relation between the Fourier transformation of the ADHM construction 
(as relevant for the finite temperature case) and the Nahm transformation, a 
crucial ingredient for our success to find explicit solutions~\cite{KvB}. 
Fig.~\ref{fig:fig1} gives a particular example for the action and 
zero-mode densities of a charge 2 caloron solution. The two dimensional 
zero-mode basis is chosen such that each zero-mode only localizes on 
one of the constituents of a given charge, showing both the zero-modes
with periodic and anti-periodic boundary conditions in the imaginary 
time direction.

This paper is organized as follows. In Sect.~\ref{sec:con} we will outline 
the construction, introduce the Green's function that is computed through 
the analogy of an impurity scattering problem, and summarize the various 
limits that can be formulated before explicitly solving for the Green's 
function.  In Sect.~\ref{sec:exact} we present the method that allows one 
to find the exact solution for the Green's function, first for the general 
case and then applied in more detail to that of topological charge 2 calorons. 
Readers only interested in the results could skip Sects.~\ref{sec:lim} and 
\ref{sec:exact}. In Sect.~\ref{sec:plots} we discuss the two classes 
of configurations in the moduli space of the charge 2 calorons and provide 
plots of the various quantities to illustrate the properties of the exact 
results. In Sect.~\ref{sec:Higgs} we discuss the relation between the 
algebraic tail of the gauge field and the zero-mode density. We end with 
some discussions. An Appendix presents a new result for the limiting 
behaviour of the action density. 

\section{Outline of the construction}\label{sec:con}

There are two steps in the construction of charge $k$ caloron solutions. 
The first step involves finding a $U(k)$ gauge field $\hat A_\mu(z)$, which
satisfies the self-duality equation, i.e. the Nahm equation~\cite{NCal}, on a 
circle parametrized by $z$, with $z$ introduced through replacing the original 
SU($n$) gauge field $A_\mu(x)$ by $A_\mu(x)-2\pi i z\delta_{0\mu}\Eins_n$. 
Although not affecting the field strength, this changes the holonomy to 
$\exp(-2\pi iz\beta)\pl$, revealing that $z$ has period $\beta^{-1}$. 
The index theorem guarantees the existence of $k$ zero-modes $\Psi(x;z)$ 
which satisfy the Dirac equation, or in the two-component Weyl form
\beq
\bar D_z\Psi(x;z)\equiv\bar\sigma_\mu D_z^\mu\Psi(x;z)\equiv\bar\sigma_\mu
(\partial_\mu+A_\mu(x)-2\pi iz\delta_{0\mu}\Eins_n)\Psi(x;z)=0.\label{eq:Dir}
\eeq
with $\bar\sigma_\mu=\sigma_\mu^\dagger=(\ein_2,-i\vec \tau)$ ($\tau_i$ are the
usual Pauli matrices). We may remove $z$ from the gauge field $A_\mu(x)$ by 
transforming the zero-mode to $\hat\Psi_z(x)\equiv\exp(-2\pi itz)\Psi(x;z)$, 
which is at the expense of making the zero-mode only periodic up to a phase 
factor, $\hat\Psi_z(t+\beta,\vec x)=\exp(-2\pi iz\beta)\hat\Psi_z(t,\vec x)$. 
In a similar way we could introduce $\vec z$ through $\Psi(x;z,\vec z)\equiv
\exp(2\pi i\vec z\cdot\vec x)\Psi(x;z)$, which replaces in \refeq{Dir} 
$A_\mu(x)$ by $A_\mu(x)-2\pi iz_\mu\Eins_n$, where $z_0\equiv z$. Assuming 
the $k$ zero-modes $\Psi^{(a)}(x;z,\vec z)$ to be orthonormal one has
\beq
\hat A^{ab}_\mu(z)=\int\Psi^{(a)}(x;z,\vec z)^\dagger\frac{\partial 
}{\partial z_\mu}\Psi^{(b)}(x;z,\vec z) d^4x,\label{eq:Ndual}
\eeq
or equivalently (demonstrating as well that $\hat A$ does not depend 
on $\vec z$)
\beq
\hat A^{ab}_0(z)=\int\Psi^{(a)}(x;z)^\dagger\frac{\partial}{\partial z}
\Psi^{(b)}(x;z) d^4x,\quad \hat A^{ab}_k(z)=2\pi i\int\Psi^{(a)}(x;z)^\dagger 
x_k\Psi^{(b)}(x;z)d^4x.
\eeq
We have shown how $\hat A_\mu(z)$ can alternatively be related to a Fourier 
transformation of the ADHM construction of instantons~\cite{ADHM}, that 
periodically repeat (up to a gauge rotation with $\pl$) in the imaginary 
time direction, so as to turn an infinite charge instanton in $R^4$ to one 
of finite charge and finite temperature. The derivation of this relation 
will not be repeated here, see Ref.~\cite{KvB,BrvB} for the details. 

The connection to the ADHM construction has been useful to simplify the 
second step in the construction of the caloron solutions, namely how to 
reconstruct the original gauge field when given a solution to the Nahm 
equation~\cite{NCal} (which is equivalent to the quadratic ADHM constraint),
\beq
\ddz\hat A_j(z)+[\hat A_0(z),\hat A_j(z)]+\half\veps_{jk\ell}[\hat A_k(z),
\hat A_\ell(z)]=2\pi i\sum_m\delta(z-\mu_m)\rho_m^{\,j},\label{eq:nahmeq}
\eeq
For convenience of notation we will henceforth use the classical scale 
invariance to set $\beta=1$. The  singularities in the Nahm equation appear 
precisely for those values where $e^{-2\pi iz}\pl$ has one of its eigenvalues 
equal to 1, at $z=\mu_i$. This is where some fermion field components become 
massless, i.e. the zero-mode becomes delocalized, whereas for generic values 
of $z$ it is exponentially localized, which has turned out to be a useful 
tool to pinpoint the constituent monopoles. 

One could apply the Nahm transformation again, introducing $\hat A_\mu(z)-2\pi
ix_\mu\Eins_k$ (with $x_0\equiv t$) and find the $n$ chiral fermion zero-modes
in the background of this gauge field. The construction is somewhat complicated 
due to the presence of the singularities, whose structure is determined from 
the matrices $\vec\rho_m$ which appear in the Nahm equation. Not all $\vec
\rho_m$ are independent, e.g. integrating and tracing the Nahm equation yields 
the conditions, $\sum_{m=1}^n\Tr\,\vec\rho_m=\vec 0$. Further constraints are 
implied~\cite{BrvB} by the fact that one may introduce (as is most easily 
seen in relation to the ADHM construction) $k$ two-component spinors 
$\zeta_a^\dagger$ in the $\bar n$ representation of SU($n$), such that 
\beq
2\pi\zeta_a^\dagger P_m\zeta_b=\sigma_0\hat S_m^{ab}-\vec\tau\cdot\vec
\rho_m^{\,ab},\label{eq:Srho}
\eeq
where $P_m$ are projections defined through $\pl=\sum_{m=1}^ne^{2\pi 
i\mu_m}P_m$ ($\hat S_m$ will appear in \refeq{VRSdef}).

\subsection{The Green's function}

However, with reference to the ADHM construction, there is great benefit
in first finding the solution for  the Green's function, $\hat f_x(z,z')\equiv
\hat g^\dagger(z)f_x(z,z')\hat g(z')$, where
\beq
\left\{-\frac{d^2}{dz^2}+V(z;\vec x)\right\}\!f_x(z,z')=4\pi^2 \ein_k
\delta(z\!-\!z'),\label{eq:green}
\eeq
with
\beqa
&& V(z;\vec x)\equiv4\pi^2\vec R^2(z;\vec x)+2\pi \sum_m\delta(z-\mu_m)S_m,
\quad S_m\equiv\hat g(\mu_m)\hat S_m\hat g^\dagger(\mu_m),\nonumber\\
&& R_j(z;\vec x)\equiv x_j-(2\pi i)^{-1}\hat g(z)\hat A_j(z)\hat g^\dagger(z),
\label{eq:VRSdef}
\eeqa
and $S_m$ playing the role of ``impurities". This is formulated in the 
gauge where first we transform $\hat A_0(z)$ to a constant (diagonal)
matrix, $2\pi i\xi_0$, as is always possible in one dimension, and then use
\beq
\hat g(z)\equiv\exp(2\pi i(\xi_0-x_0\Eins_k)z),\quad\Tr\,\xi_0=0
\eeq
(when $\Tr\,\xi_0\neq0$ it is absorbed in a shift of $x_0$) to transform 
$\hat A_0-2\pi ix_0\Eins_k$ to zero. This is at the expense of introducing 
periodicity up to a gauge transformation; although $\hat f_x(z,z')$ is 
periodic in $z$ and $z'$ with period 1 (for $\beta=1$), $f_x(z,z')$ no 
longer is\footnote{It is in this respect interesting to note that $\hat g(1)$ 
plays the role of the holonomy associated to the dual Nahm gauge field 
$\hat A_\mu(z)$.}. 

Given a solution for the Green's function, there are straightforward 
expressions for the gauge field~\cite{BrvB} (only involving the Green's 
function evaluated at the ``impurity" locations) and the fermion 
zero-modes~\cite{Us,ZM}. For the zero-mode density this gives
\beq
\hat\Psi_z^{(a)}(x)^\dagger\hat\Psi_z^{(b)}(x)=-(2\pi)^{-2}\partial_\mu^2
\hat f_x^{ab}(z,z).\label{eq:zmdens}
\eeq
In this paper we will only have need for the Green's function at $z'=z$,
which formally can be expressed as
\beq
f_x(z,z)=-4\pi^2\left((\ein_{2k}-\cF_{z})^{-1}\right)_{12},\quad\cF_{z}\equiv 
\hat g^\dagger(1)\Pexp\int_{z}^{z+1}\pmatrix{0&\ein_k\cr V(w;\vec x)&0\cr}dw,
\eeq
where the $(1,2)$ component on the right-hand side of the first identity is 
with respect to the $2\times 2$ block matrix structure. In particular this 
leads to a compact expression for the action density~\cite{KvB,BrvB}
\beq
\cS(x)\equiv-\half\tr F_{\mu\nu}^2(x)=-\half\partial_\mu^2\partial_\nu^2\log
\det\left(ie^{-\pi ix_0}(\Eins_{2k}-\cF_z)\right),\label{eq:Sdens}
\eeq
which can be shown to be independent of the choice of $z$.

The formal expression for $\cF_z$ can be made explicit by a decomposition 
into the ``impurity" contributions $T_m$ at $z=\mu_m$ and the ``propagation" 
$H_m\equiv W_m(\mu_{m+1},\mu_m)$ between $\mu_m$ and $\mu_{m+1}$. For 
$z\in(\mu_m,\mu_{m+1})$ this gives
\beq
\cF_z=W_m(z,\mu_m)T_mH_{m-1}\cdots T_2H_1T_1\hat g^\dagger(1)H_nT_n
H_{n-1}\cdots T_{m+1}H_m W_m(\mu_m,z),
\eeq
with 
\beq
T_m\equiv\pmatrix{\ein_k&0\cr 2\pi S_m&\ein_k\cr},\quad W_m(z,z')\equiv
\pmatrix{\ \ f_m^+(z)&f_m^-(z)\cr \!\ddz f^+_m(z)&\!\!\!\!\ddz f^-_m(z)\cr}
\pmatrix{\ \ f_m^+(z')&f_m^-(z')\cr \!\ddz f^+_m(z')&\!\!\!\!\ddz f^-_m(z')
\cr}^{-1}.\label{eq:TWdeg}
\eeq
The columns of the two $k\times k$ matrices $f_m^\pm(z)$, defined for 
$z\in(\mu_m,\mu_{m+1})$, form the $2k$ solutions of the homogeneous Green's 
function equation, 
\beq
\left(\frac{d^2}{dz^2}-4\pi^2\vec R^2(z;\vec x)\right)\hat v(z)=0,
\label{eq:homG}
\eeq
of which those in $f_m^+(z)$ are exponentially rising and 
those in $f_m^-(z)$ are exponentially falling. This implies~\cite{BrvB} 
$f_m^\pm(z)\to\exp\left(\pm 2\pi|\vec x|(z-\mu_m)\ein_k\right)C_m^\pm$ for 
$|\vec x|\to\infty$, in which $C_m^\pm\equiv f_m^\pm(\mu_m)$ can be arbitrary 
non-singular (to ensure a complete set of solutions) matrices. In 
Ref.~\cite{Us} we put $C_m^\pm=\Eins_k$, but here we find it convenient 
to leave this choice open. With the ``impurity" scattering problem solved,
constructing the exact solutions of the homogeneous Green's function equation
is the last step in finding analytic expressions for the higher charge
calorons.

\subsection{Limiting cases}\label{sec:lim}

Nevertheless, approximate solutions can be derived, either assuming $\vec x$ is
far removed from any core such that the gauge field has become abelian (which 
we called the far field limit, denoted by a subscript ``ff"), or assuming 
$\vec x$ and the constituents of type $m$ (belonging to the $m$-th interval) 
are well separated from all others, but not necessarily from each other (which 
was called the zero-mode limit, denoted by a subscript ``zm"). This is because 
we found~\cite{Us} that for $z\in(\mu_m,\mu_{m+1})$ the $k$ zero-modes 
$\hat\Psi^{(a)}_z(x)$ only ``see" the constituents of type $m$. For
$\mu_m\leq z\leq\mu_{m+1}$ we have
\beqa
f^\zm_x(z,z)&=&\pi\left(f_m^-(z)f_m^-(\mu_{m+1})^{-1}-f_m^+(z)
f_m^+(\mu_{m+1})^{-1}Z_{m+1}^-\right)\times\nonumber\\&&\hskip-3mm
\left(f_m^-(\mu_m)f_m^-(\mu_{m+1})^{-1}-Z_m^+f_m^+(\mu_m)
f_m^+(\mu_{m+1})^{-1}Z_{m+1}^-\right)^{-1}\times\nonumber\\&&\qquad
\left(f_m^-(\mu_m)f_m^-(z)^{-1}-Z_m^+f_m^+(\mu_m)f_m^+(z)^{-1}
\right)R_m^{-1}(z),\label{eq:zmlimit}
\eeqa
up to {\em exponential} corrections in the distance to the constituents of 
type $m'\neq m$, with
\beqa
Z_m^-\equiv\ein_k-2\Sigma_m^{-1}R_{m-1}(\mu_m),&& Z_m^+\equiv\ein_k-2
\Sigma_m^{-1}R_m(\mu_m),\label{eq:Zedefin}\\R_m(z)\equiv\half(R_m^+(z)+
R_m^-(z)),&&\Sigma_m\equiv R_m^-(\mu_m)+R_{m-1}^+(\mu_m)+S_m.\nonumber
\eeqa
and 
\beq
2\pi R_m^\pm(z)\equiv\pm\left(\ddz f_m^\pm(z)\right)f_m^\pm(z)^{-1}.
\label{eq:Rmdef}
\eeq
In the zero-mode limit $Z_m^+$ and $Z_{m+1}^-$ approach $\Eins_k$ up to 
{\em algebraic} corrections. Neglecting these contributions as well,
e.g. by sending the constituents of type $m'\neq m$ to infinity, leads 
to the so-called monopole limit (denoted by a subscript ``mon"), further 
simplifying the expression for the Green's function to 
\beq
f^{\mon}_x(z,z)=-\pi U(z,\mu_{m+1})U^{-1}_m(\mu_m,\mu_{m+1})U_m(\mu_m,z)
R_m^{-1}(z),\label{eq:zm-alg}
\eeq
with\footnote{Note that $U_m(z,z')$ satisfies the Green's function equation 
with boundary conditions $U_m(z,z)=0$ and $\frac{d}{dz}U_m(z,z')=-\frac{d}{dz'}
U_m(z,z')=2\pi R_m(z)$, for $z'\to z$.} 
\beq
U_m(z,z')\equiv f_m^+(z)f_m^+(z')^{-1}-f_m^-(z)f_m^-(z')^{-1}.\label{eq:Um} 
\eeq

In turn, from \refeq{zmlimit} one may derive the far field limit, giving up to 
{\em exponential} corrections in the distance to {\em all} constituents
\beq
f_x^\ff(\mu_m,\mu_m)=2\pi \Sigma_m^{-1},
\eeq
as is relevant for the expression of the gauge field in this limit~\cite{BrvB}.
For the zero-mode density (\refeq{zmdens}) we may use for $\mu_m<z<\mu_{m+1}$ 
($z$ strictly different from $\mu_m,\mu_{m+1}$)
\beq
f_x^\ff(z,z)=\pi R_m^{-1}(z).\label{eq:zmff}
\eeq
The discontinuity in this limit, $\pi R_m(\mu_m)\neq 2\pi\Sigma_m^{-1}\neq\pi 
R_{m-1}(\mu_m)$, arises due to the zero-mode developing a massless component 
when $z$ approaches $\mu_m$. It might seem that \refeq{zmff}, combined with 
\refeq{zmdens}, is inconsistent with an exponential decay. However, it 
turns out that~\cite{Us} 
\beq
\cV_m(\vec x)\equiv(4\pi)^{-1}\Tr\left(R_m^{-1}(z)\right) \label{eq:Vdef}
\eeq
is independent of $z$ in the $m$-th interval and harmonic everywhere (hence 
giving vanishing zero-mode density) except for some singularities in the 
cores of the constituents of type $m$. It is this feature, and our ability 
to compute $\cV_m(\vec x)$ exactly for SU(2) charge 2 calorons, that allowed 
us to make statements about the localization of the cores, without solving 
the Green's function exactly.

In Appendix A we derive the following new result for the monopole limit 
of the action density (\refeq{Sdens}). In the limit where $\vec x$ and 
the constituent locations of type $m$ are well separated from all other 
constituents, for which both the action and zero-mode densities are static, 
we find (with $U_m$ defined in \refeq{Um})
\beq
\cS^\mon(\vec x)=-\half\partial_i^2\partial_j^2\log\det\left[U_m(\mu_{m+1},
\mu_m)R_m^{-1}(\mu_m)\right],\label{eq:Szm-alg}
\eeq
up to {\em algebraic} corrections. This is a direct generalization for the 
action density of a single BPS monopole, $\cS(\vec x)=-\half\partial_i^2
\partial_j^2\log\Bigl[\sinh(2\pi\nu_m|\vec x|)/|\vec x|\Bigr]$ (located at 
the origin and with mass $8\pi^2\nu_m$). We emphasize that this result can 
be used irrespective of the distance between the constituents of the same 
type $m$. \refeq{Szm-alg} therefore gives in terms of $f_m^\pm(z)$ a closed 
form expression for the multi-monopole energy density. The same holds, when 
using \refeq{zm-alg}, for the zero-mode density (\refeq{zmdens}). For the 
caloron it involves taking the limit where all constituents of type $m'\neq 
m$ are sent to infinity, which is why this is called the monopole limit. 
We recall, that in the far field one should use~\cite{BrvB} (see also App. A)
\beq
\cS^\ff(\vec x)=-\half\partial_i^2\partial_j^2\sum_{m=1}^n\log\det\left(f_m^+
(\mu_{m+1})f_m^+(\mu_m)^{-1}R_m^{-1}(\mu_m)\Sigma_m\right).\label{eq:Sff}
\eeq

\section{Exact results}\label{sec:exact}

Finding the exact homogeneous solutions of the Green's function equation,
\refeq{homG}, closely follows Nahm's method~\cite{NCal} to construct the dual 
chiral zero-modes. The main advantage of our approach is that we need not worry 
about boundary conditions, as this is solved by the ``impurity" scattering 
formalism~\cite{BrvB}. In the following we restrict ourselves to a given 
interval $z\in(\mu_m,\mu_{m+1})$ and work in the gauge where $\hat A_0(z)
-2\pi ix_0\Eins_k=0$.

Using that $\hat A_\mu(z)$ is self-dual, a consequence of the Nahm equation, 
one easily shows that $\hat D_x^\dagger \hat D_x=-\frac{d^2}{dz^2}+4\pi^2
\vec R^{\,2}(z;\vec x)$, such that it is natural to consider the equation 
\beq
\hat D_x\hat\psi(z)=\sigma_\mu\hat D_x^\mu\hat\psi(z)=\left(\ddz-2\pi
\vec\tau\cdot\vec R(z;\vec x)\right)\hat\psi(z)=0.\label{eq:dualD}
\eeq
It follows that $\hat\psi(z)$ would be a homogeneous solution of the Green's 
function equation, albeit that $\hat\psi(z)$ is a spinor (with a chirality 
opposite to that for the zero-modes involved in the Nahm transformation, cmp. 
\refeq{Dir}). We follow Nahm~\cite{NCal} and use the ansatz $\hat\psi(z)=(
\Eins_2+\vec u(\vec x)\cdot\vec\tau)|s\rangle\otimes\hat v(z)$, where $\hat v
(z)$ is a $k$ dimensional complex vector, $\vec u(\vec x)$ is a unit vector 
that does not depend on $z$ and $|s\rangle$ is an arbitrary normalized constant
spinor (as long as it is not annihilated by $\Eins_2+\vec u(\vec x)\cdot\vec
\tau$). It then follows that $\hat v(z)=\langle s|\left(\Eins_2-\vec u(\vec 
x)\cdot\vec\tau\right)\hat\psi(z)$ satisfies \refeq{homG}.

The unit vector $\vec u(\vec x)$ is found from a complex vector $\vec y(x)$ 
which squares to 0, $\vec y(\vec x)\cdot\vec y(\vec x)=0$, implying its real 
and imaginary parts are perpendicular and of equal length ($\neq 0$ as long 
as $\vec y(\vec x)\neq\vec 0$), such that (for ease of notation the $\vec x$ 
dependence of $\vec y$ and $\vec u$ will henceforth be left implicit)
\beq
\vec u=i\vec y\times\vec y^{\,*}/\Bigl(\vec y\cdot\vec y^{\,*}\Bigr)
\label{eq:udef}
\eeq
is well defined and $\vec u\times\vec y=-i\vec y$, i.e. Re$(\vec y)$, 
Im$(\vec y)$ and $\vec u$ form an orthogonal set of vectors.
Using the ansatz for $\hat\psi(z)$ and introducing 
\beq
\Y(z)\equiv-\vec y\cdot\vec R(z;\vec x),\quad 
\U(z)\equiv-2\pi\vec u\cdot\vec R(z;\vec x),
\eeq
leads to the equations
\beq
\Y(z)\hat v(z)=0,\quad \ddz\hat v(z)=\U(z)\hat v(z)\label{eq:YU}
\eeq
for which the first one can only have a solution provided $\det \Y(z)=0$.

It is the great beauty of Nahm's formalism that $\det \Y(z)$ is a conserved 
quantity. That is, $\ddz\hat A_j(z)=-\half\veps_{jk\ell}[\hat A_k(z),\hat 
A_\ell(z)]$ (the Nahm equation \refeq{nahmeq} restricted to an interval and 
in the gauge where $\hat A_0=0$) implies $\ddz\det \Y(z)=0$ for any choice 
of $\vec x$ and $\vec y$ on a null-cone in $C^3$ (or rather in $CP^2$ since 
we may rescale $\vec y$ with a non-zero complex factor without changing the 
equations). The conserved quantities are generated by the symmetric traceless 
monomials, $M_{i_1i_2\cdots i_\ell}$, built from $\Tr\left(\hat A_{i_1}(z)
\hat A_{i_2}(z)\cdots\hat A_{i_\ell}(z)\right)$ with $\ell$ arbitrary, as 
one readily verifies. For example, $\Tr\,\hat A_i(z)$ is constant and defines 
the center of mass. A natural way to project on the traceless symmetric 
monomials is precisely through introducing $\vec y\in CP^2$ on a null-cone, 
forming $y_{i_1}y_{i_2}\cdots y_{i_\ell}M_{i_1i_2\cdots i_\ell}=y_{i_1}
y_{i_2}\cdots y_{i_\ell}\Tr\left(\hat A_{i_1}(z)\hat A_{i_2}(z)\cdots\hat 
A_{i_\ell}(z)\right)$. It is interesting to note that $x_{i_1}x_{i_2}\cdots 
x_{i_\ell}M_{i_1i_2\cdots i_\ell}|\vec x|^{-\ell}$ is always a spherical 
harmonic of order $\ell$, used in Ref.~\cite{Us} to show through the 
multipole expansion of $\Tr\left(R_m^{-1}(z)\right)$ that it is conserved 
and harmonic, except for singularities in the core of the constituents. 

Once it is established that $\det \Y(z)$ is independent of $z$ for any choice 
of $\vec x$, we can look for its zeros. Using the null-cone parametrization 
$\vec y=(\half(1-\zeta^2),-\frac{i}{2}(1+\zeta^2),\zeta)$, and the fact that 
the matrix $\Y(z)$ is $k$ dimensional, we obtain a polynomial equation in 
$\zeta$ of order $2k$ and hence there are for generic $\vec x$ exactly $2k$ 
solutions. It is useful to note that these solutions come in complex conjugate 
pairs, where the symmetry $\vec y\to\vec y^{\,*}$ implies $\vec u\to-\vec u$ 
and $\zeta\to-1/\zeta^*$ (giving $\vec y^{\,*}$ up to a multiple, equivalent 
to $\vec y^{\,*}$ in $CP^2$).

Given a particular zero $\vec y$, we may conveniently write a vector in the 
kernel of $\Y(z)$ as~\cite{NAlg} $\hat v_a(z)=\hat\phi(z)(\adj\Y(z))_{ac}$ 
for a fixed choice of $c$, where $\adj\Y(z)$ is the matrix formed by the 
minors of $\Y(z)$. The Nahm equation is easily seen to imply $\ddz\Y(z)=
[\U(z),\Y(z)]$ for {\em any} choice of $\vec y$ on the null-cone. From 
this one derives that\footnote{Assume first that $\vec y$ is such that 
$\det\Y\neq 0$, in which case $\adj\Y=\Y^{-1}\det\Y$ and therefore 
$\ddz\adj\Y=-\Y^{-1}[\U,\Y]\Y^{-1}\det\Y+\Y^{-1}\ddz\det\Y=[\U,\adj\Y]
+\Y^{-1}\Tr(\Y^{-1}\ddz\Y)=[\U,\adj\Y]$. Observe that $\adj\Y$ is analytic 
in $\vec y$, such that the result is valid also when $\vec y$ leads to 
$\det\Y=0$.} $\ddz\adj\Y(z)=[\U(z),\adj\Y(z)]$. Substituting $\hat v_a(z)
=\hat\phi(z)(\adj\Y(z))_{ac}$ into \refeq{YU} gives 
\beq 
\frac{d\hat\phi(z)}{dz}(\adj\Y(z))_{ac}=\hat\phi(z)\Bigl(\adj\Y(z)\U(z)
\Bigr)_{ac}.
\eeq
Using the fact that $\U(z)=-iu_j\hat g(z)\hat A_j(z)\hat g^\dagger(z)-2\pi u_j
x_j$, we get
\beq
\hat\phi(z)=\frac{\exp\Bigl(\hat\mu(z)-2\pi z\vec u\cdot\vec x\Bigr)}{\sqrt{
(\adj\Y(z))_{ac}}},\quad\frac{d\hat\mu(z)}{dz}=-\frac{i\Bigl\{\adj\Y(z),
u_j\hat g(z)\hat A_j(z)\hat g^\dagger(z)\Bigl\}_{ac}}{2(\adj\Y(z))_{ac}},
\label{eq:mu}
\eeq
where the equation for $\hat\mu(z)$ is the same for any value of $a$ (it 
may depend on the value of $c$). This is useful for studying the asymptotic 
behaviour of the solution. For large $|\vec x|$, $\det\Y(z)=0$ implies that 
$\vec y\cdot\vec x\to0$, such that (cmp. \refeq{udef}) $\vec u\to\pm\vec x/
|\vec x|$. We may use the symmetry $\vec u\to-\vec u$ to guarantee that there 
are $k$ zeros $\vec y^{\,(b)}$ with the sign of $\vec u^{\,(b)}\!\cdot\vec x$ 
negative, leading to solutions that rise as $\exp(2\pi z|\vec x|)$. It then 
follows that the $k$ zeros $\vec y^{\,(b+k)}=\vec y^{\,(b)*}$ give $\vec u^{
\,(b+k)}=-\vec u^{\,(b)}$, leading to the solutions that decay as $\exp(-2\pi 
z|\vec x|)$ for large $|\vec x|$. Hence we may put $f_{ab}^+(z)=\hat v_a^{(b)}
(z)$ and $f_{ab}^-(z)=\hat v_a^{(b+k)}(z)$. Defining 
\beq
\hat m_{ab}^+(z)=-\left(\adj \vec y^{\,(b)}\!\cdot\vec R(z;\vec x)
\right)_{ac},\quad 
\hat m_{ab}^-(z)=-\left(\adj \vec y^{\,(b)*}\!\!\cdot\vec R(z;\vec x)
\right)_{ac},\label{eq:mdef}
\eeq
which are {\em algebraic} in $\hat A_j(z)$ and $x_j$, we find
\beq
f^+_{ab}(z)=\hat m^+_{ab}(z)\hat\phi^{(b)}(z),\quad
f^-_{ab}(z)=\hat m^-_{ab}(z)\hat\phi^{(b+k)}(z),
\eeq
where $\hat\phi(z)$ contains the exponential dependencies, and thus seemingly
all the information about the cores of the constituents. To make this more 
precise we compute $R^\pm(z)$, see \refeq{Rmdef}, using that \refeq{YU} 
implies (with $\vec u^{\,(b+k)}=-\vec u^{\,(b)}$) $\ddz f^\pm_{ab}(z)=\mp
2\pi\sum_{d=1}^k \vec u^{\,(b)}\cdot\vec R_{ad}(z;\vec x)f_{db}^\pm(z)$, 
finding that the factors $\hat\phi^{(b)}(z)$ {\em drop out} 
\beq
R_{ad}^\pm(z)=-\sum_{b,e=1}^k\vec u^{\,(b)}\cdot\vec R_{ae}
(z;\vec x)\hat m_{eb}^\pm(z)(\hat m^\pm(z)^{-1})_{bd}.\label{eq:alg}
\eeq
This proves that $R_m^\pm(z)$ is purely algebraic in $\hat A_j(z)$ and $x_j$, 
as are $\Sigma_m$ and $R_m(z)$, which determine the far field limit for the 
zero-mode density and the gauge field. 

One might wonder how, given that the $\hat\psi^{(b)}(z)$ are of the ``wrong" 
chirality in the context of the Nahm transformation, one could use these 
results to reconstruct the gauge field for magnetic monopoles where the 
relation to the ADHM construction is not readily available. For this one 
observes that the columns of $\hat\psi^{(b)}(z)$ can be used to form a 
$2k\times 2k$ matrix $w(z)$. Using that $\hat D_x w(z)=0$, one finds 
$\hat D_x^\dagger\left(w^\dagger(z)^{-1}\right)=0$. Thus the columns of 
$w^\dagger(z)^{-1}$ give $2k$ independent solutions for each interval, 
from which $n$ normalizable solutions $\hat\Psi^{(p)}(z;\vec x)$ should 
remain after imposing the appropriate boundary (cq. matching) conditions. 
These are then used in Nahm's original construction to compute the gauge 
field (cmp. \refeq{Ndual})
\beq
A^{pq}_\mu(x)=\int\hat\Psi^{(p)}(z;x_0,\vec x)^\dagger\frac{\partial 
}{\partial x_\mu}\hat\Psi^{(q)}(z;x_0,\vec x) dz,
\eeq
where $\hat\Psi^{(p)}(z;x_0,\vec x)\equiv\hat g^\dagger(z)\hat\Psi^{(p)}
(z;\vec x)$. 

There seems to be considerable advantage in using the Green's function (Fourier
transformed ADHM) method, since it can solve the matching conditions without 
relying on the availability of exact solutions for the normalizable dual 
zero-modes.  To go beyond the approximations discussed in Sect.~\ref{sec:lim} 
and resolve the constituent cores we need to solve for $\hat\mu(z)$. This can 
not always be done in closed form, but it is given by an explicit integral 
which can be performed  numerically when required. For charge 2 monopoles 
Panagopoulos~\cite{Hari} was, however, able to find the exact integral. We 
can use the same ingredients for the caloron case and explicitly solve for 
the Green's function in the case of charge 2 calorons.

\subsection{Analytic expressions for charge 2}\label{sec:ch2}

For charge 2 the number of invariants associated to the conserved quantities 
of the Nahm equation is 8, of which $\Tr\,\hat A_j(z)\equiv 4\pi ia_j$ are 
related to the center of mass for the constituents of given magnetic charge, 
coming from the interval under consideration. Assuming now that $\hat A_j(z)$ 
is traceless, 5 invariants remain, given in terms of the symmetric traceless 
tensor 
\beq
M_{ij}\equiv -\half\left(\Tr(\hat A_i(z)\hat A_j(z))
-\third\delta_{ij}\Tr(\hat A_k^2(z))\right).
\eeq
Three of its parameters are associated to the rotation $\cR$ which
diagonalizes the $3\times 3$ matrix, $M=\cR\diag(c_1,c_2,c_3)\cR^t$,
where $\cR$ is fixed by requiring $c_2\leq c_1\leq c_3$. The $c_i$
add to zero and can be expressed in terms of the so-called scale ($D$) 
and shape ($\k$) parameters,
\beq
c_1=D^2\frac{1-2\k^2}{12},\quad c_2=D^2
\frac{\k^2-2}{12},\quad c_3=D^2\frac{1+\k^2}{12}.
\eeq

The Nahm equation for the case of charge 2 can be solved completely in 
terms of Jacobi elliptic functions~\cite{NahmM,BrDa}, which was summarized
in Ref.~\cite{Us} 
\beq
\hat g(z)\hat A_j(z;\vec a,\cR,h,D,\k)\hat g^\dagger(z)\equiv 2\pi i a_j\ein_2
+\half i D\cR_{jb}f_{b}\left(D(z-z_0)\right)h^\dagger\tau_bh,\label{eq:Jnahm}
\eeq
where $h$ is a global gauge parameter and
\beq
f_1(z)\equiv\frac{\k'}{\cn_\k(z)},\quad f_2(z)\equiv\frac{\k'\sn_\k(z)}{\cn_\k
(z)},\quad f_3(z)\equiv\frac{\dn_\k(z)}{\cn_\k(z)},\quad\k'\equiv\sqrt{1-\k^2}.
\label{eq:fs}
\eeq 
with\footnote{$\vphi(z)$ implicitly defined by the elliptic integral of 
the first kind $z=\int^{\vphi(z)}_01/\sqrt{1-\k^2\sin^2\theta}~d\theta$.}
$\sn_\k(z)=\sin(\vphi(z))$, $\cn_\k(z)=\cos(\vphi(z))$ and $\dn_\k(z)=\sqrt{
1-\k^2\sn_\k^2(z)}$ the standard Jacobi elliptic functions. This does not yet 
address the matching of $\hat A_j(z)$ on the different intervals, where some 
difference between the monopole and caloron application appears. For the 
caloron, apart from the axially symmetric solutions constructed in 
Ref.~\cite{BrvB}, we found two sets of non-trivial solutions that interpolate 
between overlapping and well-separated constituents. It is for these classes 
of solutions that we will resolve the cores, when constituents overlap and 
the non-linearity plays an important role. 

The next step in the construction is finding the zeros $\vec y$ of $\det\Y(z)$.
One has
\beq
\det\Y(z)=y_iy_j(x_ix_j-\third\vec x^{\,2}\delta_{ij}-(2\pi)^{-2} M_{ij}),
\eeq
where we used $\vec y^{\,2}=0$. Substituting a parametrization for this
null-cone in $CP^2$, e.g. $\vec y=(\half(1-\zeta^2),-\frac{i}{2}(1+\zeta^2),
\zeta)$, gives a 4th order polynomial. However, for finding the 4 solutions 
we find it in this case more convenient to first diagonalize the matrix 
$x_ix_j-\third\vec x^{\,2}\delta_{ij}-(2\pi)^{-2} M_{ij}=\cO_{ik}\lambda_k
\cO_{jk}$. Introducing $\vec y^{\,\prime}=\cO^t\vec y$, the equation 
for the zeros reduces to $(y^\prime_1)^2\lambda_1+(y^\prime_2)^2\lambda_2+
(y^\prime_3)^2\lambda_3=0$, which in addition to the null-cone condition, 
$(y^\prime_1)^2+(y^\prime_2)^2+ (y^\prime_3)^2=0$, is now easily solved by 
\beq
\vec y^{\,(a)\,\prime}\!=\left(\sqrt{\lambda_2-\lambda_3},(-1)^{a+1}
\sqrt{\lambda_3-\lambda_1},i\sqrt{\lambda_2-\lambda_1}\right),\quad
\vec y^{\,(a+2)\,\prime}\!=(\vec y^{\,(a)\,\prime})^*,\  a=1,2,\label{eq:ys}
\eeq
where we fixed (for generic $\vec x$) the diagonalizing rotation $\cO$ 
by ordering $\lambda_1\leq\lambda_3\leq\lambda_2$. Using \refeq{udef} we find
for $\vec u^{\,\prime}\equiv\cO^t\vec u$
\beq
\vec u^{\,(a)\,\prime}\!=-\left((-1)^a\frac{\sqrt{\lambda_3-\lambda_1}}{\sqrt{
\lambda_2-\lambda_1}},\frac{\sqrt{\lambda_2-\lambda_3}}{\sqrt{\lambda_2- 
\lambda_1}},0\right),\quad\vec u^{\,(a+2)\,\prime}\!=-\vec u^{\,(a)\,\prime},
\  a=1,2.
\eeq
One easily checks that $\vec u^{\,(1,2)}$ and $\vec u^{\,(3,4)}$ give rise to
respectively the exponentially rising and falling solutions.

It is also instructive to give the explicit expressions for the matrices
$\hat m^\pm(z)$ in \refeq{mdef} (choosing $c=2$). We note that for 
a $2\times 2$ matrix $\adj\Y=(\Tr\,\Y)\Eins_2-\Y$, and without loss in 
generality\footnote{We may change $z_0$, $\vec a$, $h$, $\cR$ and $D$ by 
resp. translations, (gauge) rotations, and suitable rescalings.} 
we take $z_0=0$, $\vec a=\vec 0$, $\cR=\Eins_3$, $D=1$ and $h=\Eins_2$ in 
\refeq{Jnahm}, such that
\beqa
\hat m^+(z)=-\frac{1}{4\pi}\pmatrix{y_1^{(1)}f_1(z)-iy_2^{(1)}
f_2(z)&y_1^{(2)}f_1(z)-iy_2^{(2)}f_2(z)\cr
4\pi\vec x\cdot\vec y^{\,(1)}-y_3^{(1)}f_3(z)&
4\pi\vec x\cdot\vec y^{\,(2)}-y_3^{(2)}f_3(z)\cr},
\nonumber\\&&\nonumber\\
\hat m^-(z)=-\frac{1}{4\pi}\pmatrix{y_1^{(3)}f_1(z)-iy_2^{(3)}
f_2(z)&y_1^{(4)}f_1(z)-iy_2^{(4)}f_2(z)\cr
4\pi\vec x\cdot\vec y^{\,(3)}-y_3^{(3)}f_3(z)&
4\pi\vec x\cdot\vec y^{\,(4)}-y_3^{(4)}f_3(z)\cr}.
\eeqa
We checked that \refeq{Vdef}, $\cV(\vec x)=(2\pi)^{-1}\Tr(R^+(z)+R^-(z))^{-1}$,
evaluated using \refeq{alg} is independent of $z$ and agrees with the result 
derived in Ref.~\cite{Us}.

We next address solving $\refeq{mu}$, which for charge 2 can be written 
as\footnote{Using that $\half\{\hat A_i,\hat A_j\}y_iu_j=\half\Eins_2\Tr(
\hat A_i\hat A_j)y_iu_j=-M_{ij}y_iu_j\Eins_2$, since $\vec u\cdot\vec y=0$.}
\beq
\frac{d\hat\mu(z)}{dz}=\frac{M_{ij}u_iy_j\delta^{ac}
-2\pi i(\vec x\cdot\vec y)u_i\hat A_i^{ac}(z)}{2\pi
(\vec x\cdot\vec y)\delta^{ac}-iy_i\hat A_i^{ac}(z)}.
\eeq
For the same choice of parameters in \refeq{Jnahm} as above, $z_0=0$, $\vec a
=\vec 0$, $\cR=\Eins_3$, $D=1$ and $h=\Eins_2$, this gives (with 
$a=1$ and $c=2$)
\beq
\frac{d\hat\mu(z)}{dz}=2\pi(\vec x\cdot\vec y)\frac{u_1f_1(z)
-iu_2f_2(z)}{y_1f_1(z)-iy_2f_2(z)}.\label{eq:mup}
\eeq
Although the dependence on $\vec x$ is complicated, the integral over 
$z$ turns out to be manageable (as was observed before in the context of
charge 2 monopoles, although we here choose not to express the solution 
in terms of theta functions~\cite{Hari}). To solve the equation we first
rewrite the right-hand side of \refeq{mup} using the fact that $\vec u
\times\vec y=-i\vec y$ (cmp. \refeq{udef}), 
\beqa
\frac{u_1f_1(z)-iu_2f_2(z)}{y_1f_1(z)-iy_2f_2(z)}&=&\frac{(u_3y_1+iy_2)
f_1(z)-i(u_3y_2-iy_1)f_2(z)}{y_3\left(y_1f_1(z)-iy_2f_2(z)\right)}
\nonumber\\&=&\frac{f_1(z)f_2(z)y_3^2+4iy_1y_2(\k')^2}{y_3\left
(16\pi^2(\vec x\cdot\vec y)^2-y_3^2f_3^2(z\right))}+\frac{u_3}{y_3}.\nonumber
\eeqa
In the last identity we used that $\vec y^{\,2}=0$, $f_1^2(z)-f_2^2(z)=1-\k^2
=(\k')^2$ and the fact that $\det\Y(z)=0$ implies $(y_1f_1(z)-iy_2f_2(z))(y_1
f_1(z)+iy_2f_2(z))=16\pi^2(\vec x\cdot\vec y)^2-y_3^2f_3^2(z)$. With $\ddz 
f_3(z)=f_1(z)f_2(z)$ we can now integrate \refeq{mup}, 
\beqa
\hat\mu(z)&=&2\pi z(\vec x\cdot\vec y)\frac{u_3}{y_3}+\quart\log\left(
\frac{4\pi\vec x\cdot\vec y+f_3(z)y_3}{4\pi\vec x\cdot\vec y-f_3(z)y_3}
\right)+i\frac{\sign(z)(\k')^2}{2\pi(\vec x\cdot\vec y)}\frac{y_1y_2}{4
y_3}I(z),\nonumber\\I(z)&\equiv&\Pi_\k(f_3^{-1}(z),n)-\Pi_\k(1,n)+|z|,
\quad n\equiv\frac{(4\pi\vec x\cdot\vec y)^2}{y^2_3},
\eeqa
up to an irrelevant constant, where $\Pi_\k(s,n)$ is the elliptic integral of 
the third kind\footnote{More commonly the elliptic integral of the third kind 
is defined as $\Pi(n;\vphi,\k)=\Pi_\k(\sin\vphi,n)$. Note that $I(z)$ can 
alternatively be written as $\int_1^{f_3(z)}(1-t^2/n)^{-1}(t^2-\k^2)^{-\half}
(t^2-1)^{-\half}dt$, from which it follows that $\ddz I(z)=\left((1-n f_3^2(z))
|f_1(z)f_2(z)|\right)^{-1}\ddz f_3(z)=\sign(z)(1-nf_3^2(z))^{-1}$, using $f_3^2
(z)-1=f_2^2(z)$, $f_3^2(z)-\k^2=f_1^2(z)$ and $\ddz f_3(z)=f_1(z)f_2(z)$.}, 
\beq
\Pi_\k(s,n)\equiv\int_0^s\!\frac{dt}{(1-nt^2)\sqrt{(1-\k^2t^2)(1-t^2)}}.
\eeq

We now combine these ingredients to give in terms of $\vec y$ the exact form 
for the homogeneous solution of the Green's function equation, 
\beq
\hat v_a(z)=\hat\phi(z)(\adj\Y(z))_{a2}=\exp\left(\hat\mu(z)-2\pi z\vec u
\cdot\vec x\right)\frac{(\adj\Y(z))_{a2}}{\sqrt{(\adj\Y(z))_{12}}}
\eeq
or putting in all the relevant expressions
\beqa
\hat v_a(z)&=&\exp\left(i\frac{z}{y_3}\left[2\pi(\vec y\times\vec x)_3+
\frac{y_1y_2(\k')^2}{8\pi(\vec x\cdot\vec y)|z|}\left(|z|+\Pi_\k(f_3^{-1}
(z),n)-\Pi_k(1,n)\right)\right]\right)\times\nonumber\\ &&(4\pi)^{-\half}
\Bigl(-y_1f_1(z)-(-1)^aiy_2f_2(z)\Bigr)^\half\left(\frac{4\pi\vec x\cdot
\vec y-(-1)^ay_3f_3(z)}{4\pi\vec x\cdot\vec y+(-1)^ay_3f_3(z)}\right)^\quart.
\eeqa
Substituting $\vec y=\vec y^{(b)}=\cO^t\vec y^{\,(b)\,\prime}$, with 
$\vec y^{\,(b)\,\prime}$ as defined in \refeq{ys} gives $f_{ab}^+(z)=
\hat v_a^{(b)}(z)$ and $f_{ab}^-(z)=\hat v_a^{(b+2)}(z)$, and thereby the 
Green's function, once we specify the parameters involved in the solutions 
to the Nahm equation.

\section{Action and zero-mode density plots}\label{sec:plots}

The discontinuities in $\hat A_j(z)$ at $z=\mu_1$ and $z=\mu_2$ implied by the 
Nahm equation, \refeq{nahmeq}, impose constraints which are (like the quadratic
ADHM constraint) in general difficult to solve. Work is in progress to describe 
the full parameter space for SU(2) and charge 2, but in Ref.~\cite{Us} we did 
find two non-trivial parametrizations for which we illustrate here in a number 
of figures how the full caloron solutions look like, using the exact Green's 
function as constructed in the previous section. Taking advantage of the
possibility to work with arbitrary arithmetic precision the programme Maple 
has been used for these calculations. The configurations are formulated in 
terms of the two intervals $z\in[\mu_1,\mu_2]$ and $z\in[\mu_2,1+\mu_1]$, 
each associated with two constituent monopoles of equal magnetic charge, 
but opposite in sign from one interval to the next. Apart from a shift and 
(gauge) orientation, the configuration is described by a shape ($\k$) and 
scale ($D$) parameter (see Sect.~\ref{sec:ch2}), for simplicity assumed to 
be the same on both intervals. We also take all constituents to be of equal 
mass, $\nu_{1,2}=\half$ ($\mu_2=-\mu_1=\quart$), most relevant for the 
confined phase with $\tr\,\pl=0$.

The two periodic and two anti-periodic chiral fermion zero-modes each have 
support on oppositely charged constituents (see Fig.~\ref{fig:fig1}) and in 
the far field limit it was found that the zero-mode density (summed over 
the two zero-modes implied by the index theorem) is described by a disc 
singularity, bounded by an ellipse with semi-major axis $D/4\pi$ and 
eccentricity $(1-\k')/(1+\k')$ (where $\k'=\sqrt{1-\k^2}$). This revealed 
that the core of a cluster of like-charged constituents is in general 
extended, unless the individual constituents are well separated. The far 
field only describes the (algebraic) abelian component of the gauge field 
and to resolve the structure of the core we need to determine the full 
non-abelian structure.

\begin{figure}[htb]
\vspace{6.5cm}
\includegraphics{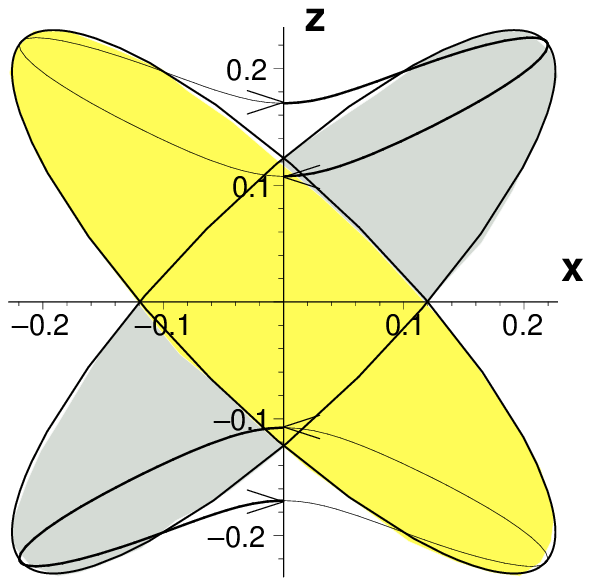}   
\includegraphics{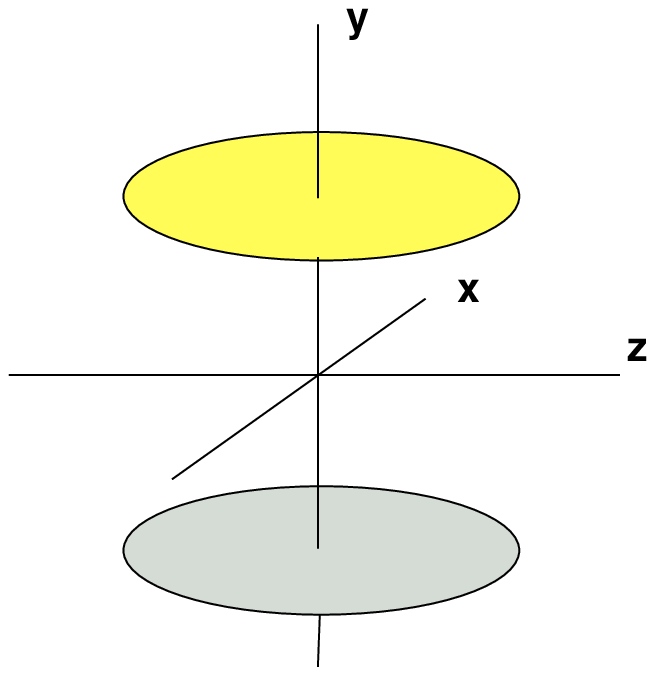}   
\caption{An example to illustrate the location of the disc singularities 
(light and dark shaded according to magnetic charge) for a rectangular 
(left) and crossed (right) configuration. The latter is shown at $\alpha=
-\pi/2$, for $d=\pi/32$ and $\theta=\pi/4$ ($\k=0.962$, $D=3.894$ and $\vphi
=-\pi/4$). The curves indicate the would-be constituent locations at fixed 
$d$ and $\theta$, varying $\alpha$ from $-\pi$ (to which the arrows point) 
to 0. For $\alpha=-\pi$ and 0 the discs collapse to lines ($\k=1$) with no 
singularities remaining, except at the endpoints.}\label{fig:conloc}
\end{figure}

For the first parametrization (called ``rectangular") the two discs are
parallel and separated in height by a distance $d$.  The configuration 
is charaterized by the two extremal points along the major axis of the 
disc, also called {\em would-be} constituent locations\footnote{In their 
immediate neighbourhood the action density is maximal; they are the 
constituent locations in the point-like limit, $\k\to1$.},
\beq
\vec y_m^{\,(j)}=\left(0\,,\,\half(-1)^md\,,\,(-1)^j(4\pi)^{-1}D\right),\quad
2\pi d=Df_2(\quart D),
\eeq
up to an overall shift and orientation (the definition of $f_2(z)$, which
involves $\k$, can be found in \refeq{fs}). A typical example is shown in 
Fig.~\ref{fig:conloc} (left). Apart from $\k$ and $D$, the parameters that 
enter \refeq{Jnahm} for the $m$-th interval are $\vec a=(0,\half(-1)^md,0)$, 
$\cR=\Eins_3$, $h=\Eins_2$ and $z_0=\quart(1+(-1)^m)$. In all cases discussed 
here we have also put $\xi_0=0$. One verifies that the discontinuities of 
$\hat A_j(z)$ at $z=\mu_m$ are given by $2\pi i\rho_m^j$ with appropriately 
chosen $\zeta_a$ (cmp. \refeq{Srho}), as discussed in detail in Ref.~\cite{Us}.

\begin{figure}[htb]
\vspace{7.1cm} 
\includegraphics{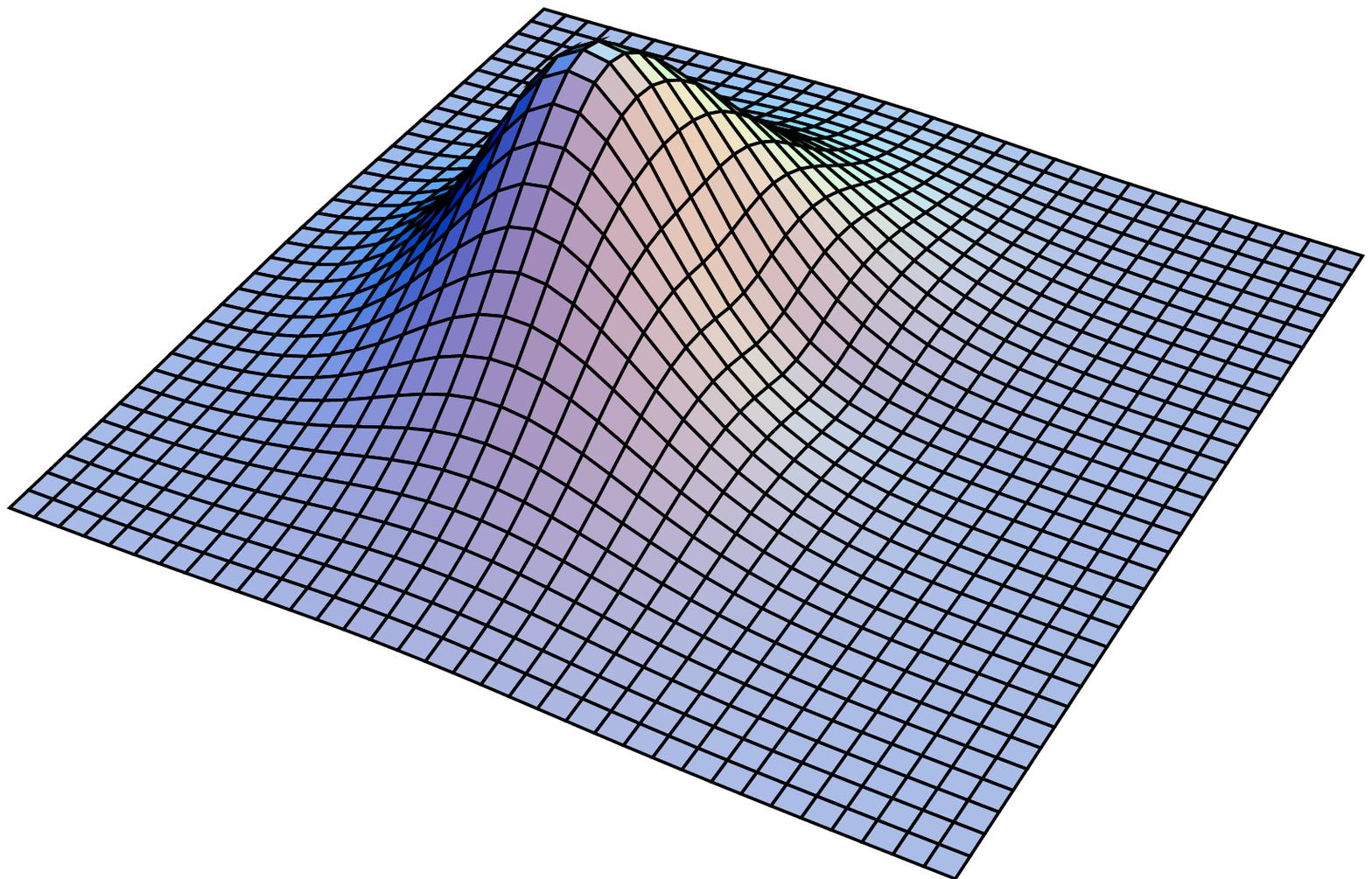}
\includegraphics{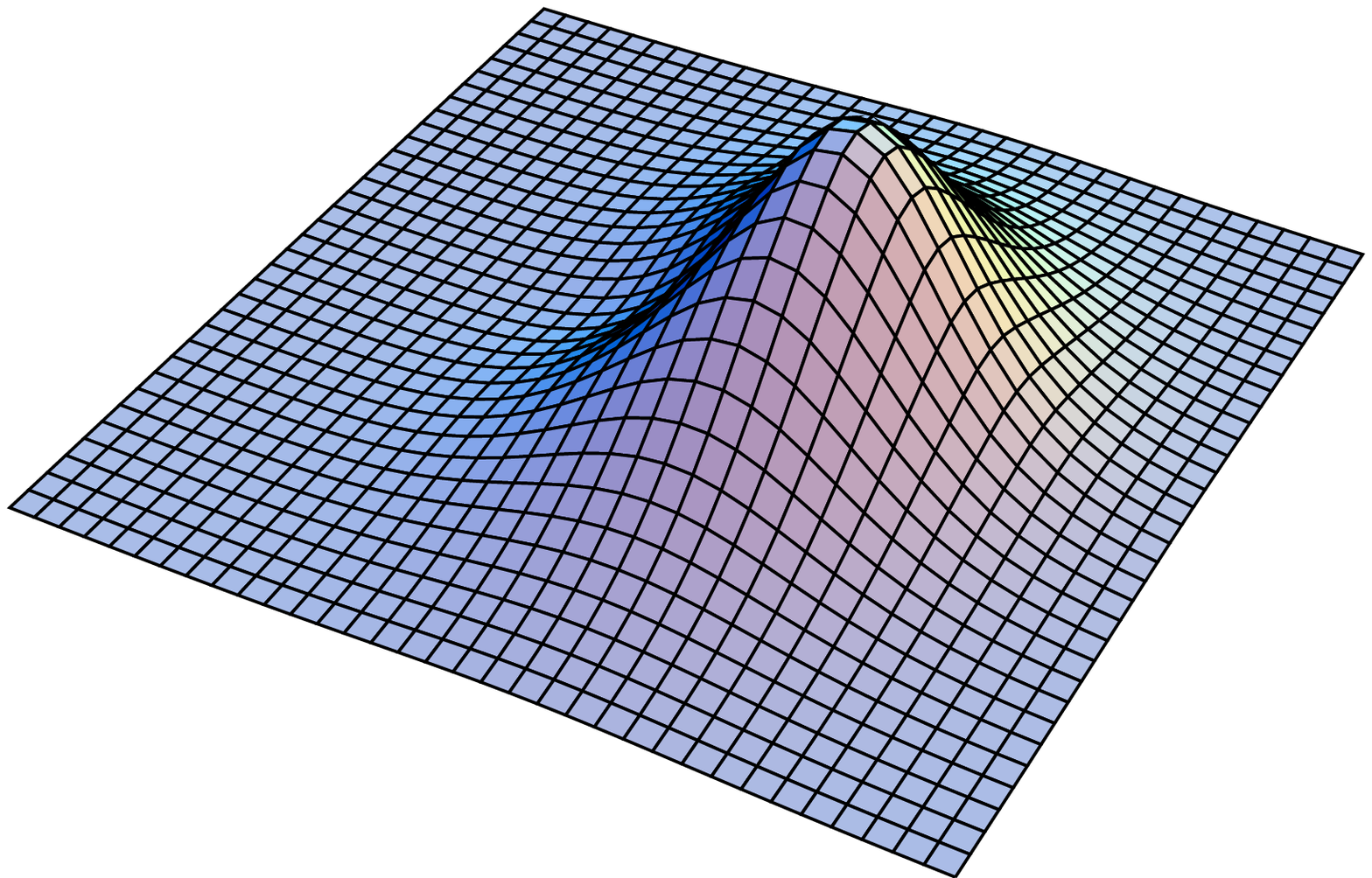}
\includegraphics{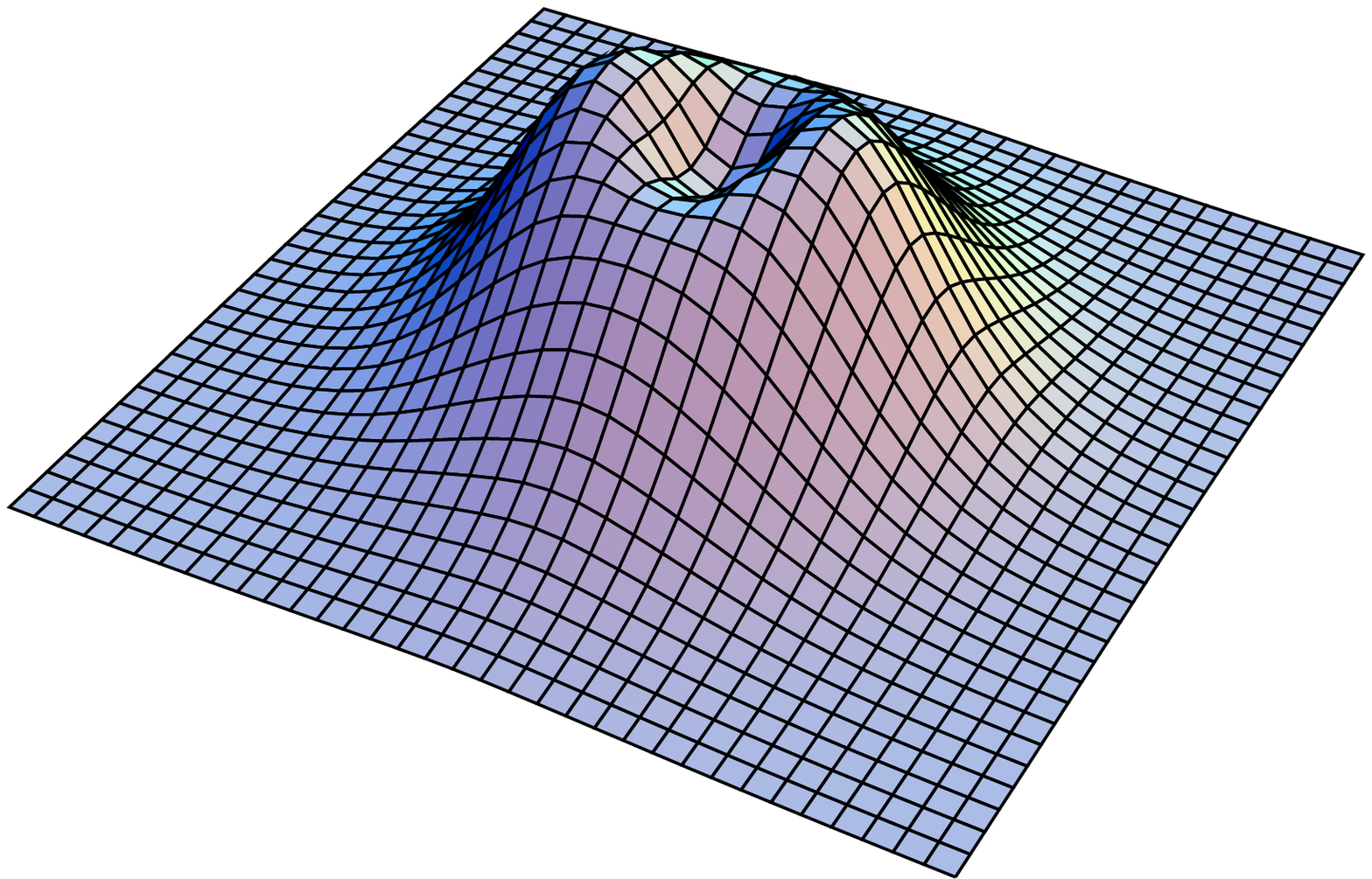}
\caption{Shown is the energy density (middle) for $\mu_2=0.25$, $\k=0.570$ and 
$D=6.915$, in the monopole limit $d\to\infty$ and in the $x$-$z$ plane through 
one of the discs for the rectangular configuration, see Fig.~\ref{fig:conloc} 
(left). On a scale enhanced by a factor $4\pi^2$ are shown the densities for 
the two monopole zero-modes (left and right).}\label{fig:monlim} 
\end{figure}

For the second parametrization (called ``crossed") the two discs are coplanar
and intersect, see Fig.~\ref{fig:conloc} (right) for a typical example. Their 
relative orientations can vary between perpendicular and coinciding (for which 
$\k$ is forced to 1). Here we choose for the parameters in \refeq{Jnahm} $h$ 
and $\cR$ to be non-trivial (isospin) rotations around the $y$-axis with 
angles $(-1)^m\theta$, resp. $(-1)^m\vphi$, and $\vec a=(0,0,-\half(-1)^md
\cos\alpha)$, whereas $z_0$ and $\xi_0$ are as in the rectangular case. The 
would-be constituent locations are now given (up to an overall shift and 
orientation) by
\beq
\vec y_m^{\,(j)}=\left((-1)^j(4\pi)^{-1}D\sin\vphi\,,\,0\,,\,(-1)^{m+j}
(4\pi)^{-1}D\cos\vphi-\half(-1)^md\cos\alpha\right),
\eeq
where $\pm\vphi$ conveniently gives the orientation of each of the two 
discs with respect to the $z$-axis. The angle $\alpha$ originates from the 
definition of $\zeta_a$, which through \refeq{Srho} determines the 
discontinuity of $\hat A_j(z)$. To ensure the proper matching, the 
following three equations need to be satisfied~\cite{Us} 
\beq
D\sin(\theta\pm\vphi)\left[f_3(\quart D)\pm f_1(\quart D)\right]=8\pi 
d(1\pm\sin\alpha),\quad Df_2(\quart D)+8\pi d\sin\alpha=0.
\eeq
which determine $\vphi$, $\k$ and $D$ for given $\alpha$, $\theta$ and 
$d$. Fig.~\ref{fig:conloc} illustrates that for these crossed configurations 
the two discs always overlap, unlike for the rectangular case. This formed 
an important motivation for the present study, so as to determine in how far 
the overlapping discs would affect the behaviour in the core. 

\begin{figure}[htb]
\vspace{9.3cm} 
\includegraphics{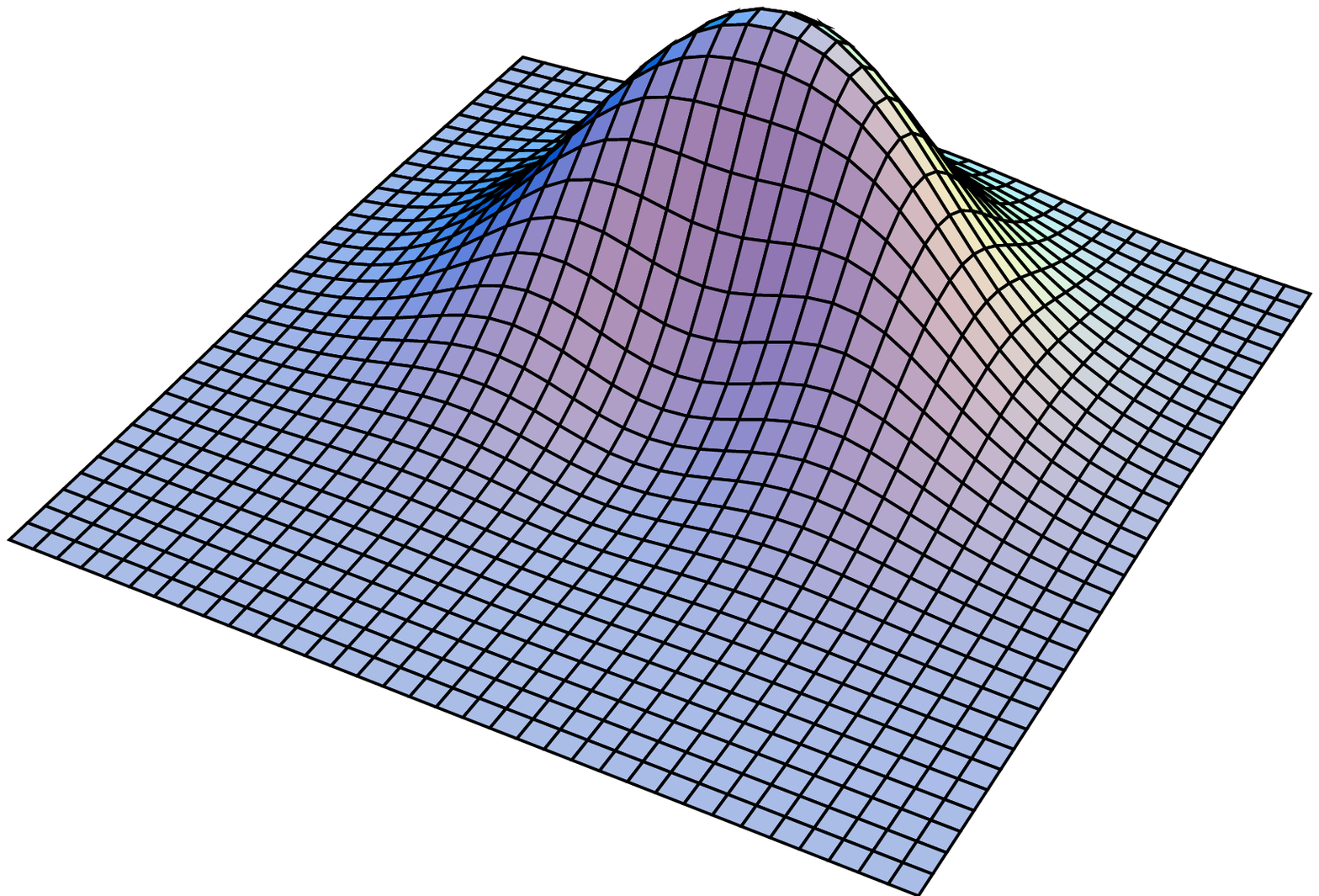}
\includegraphics{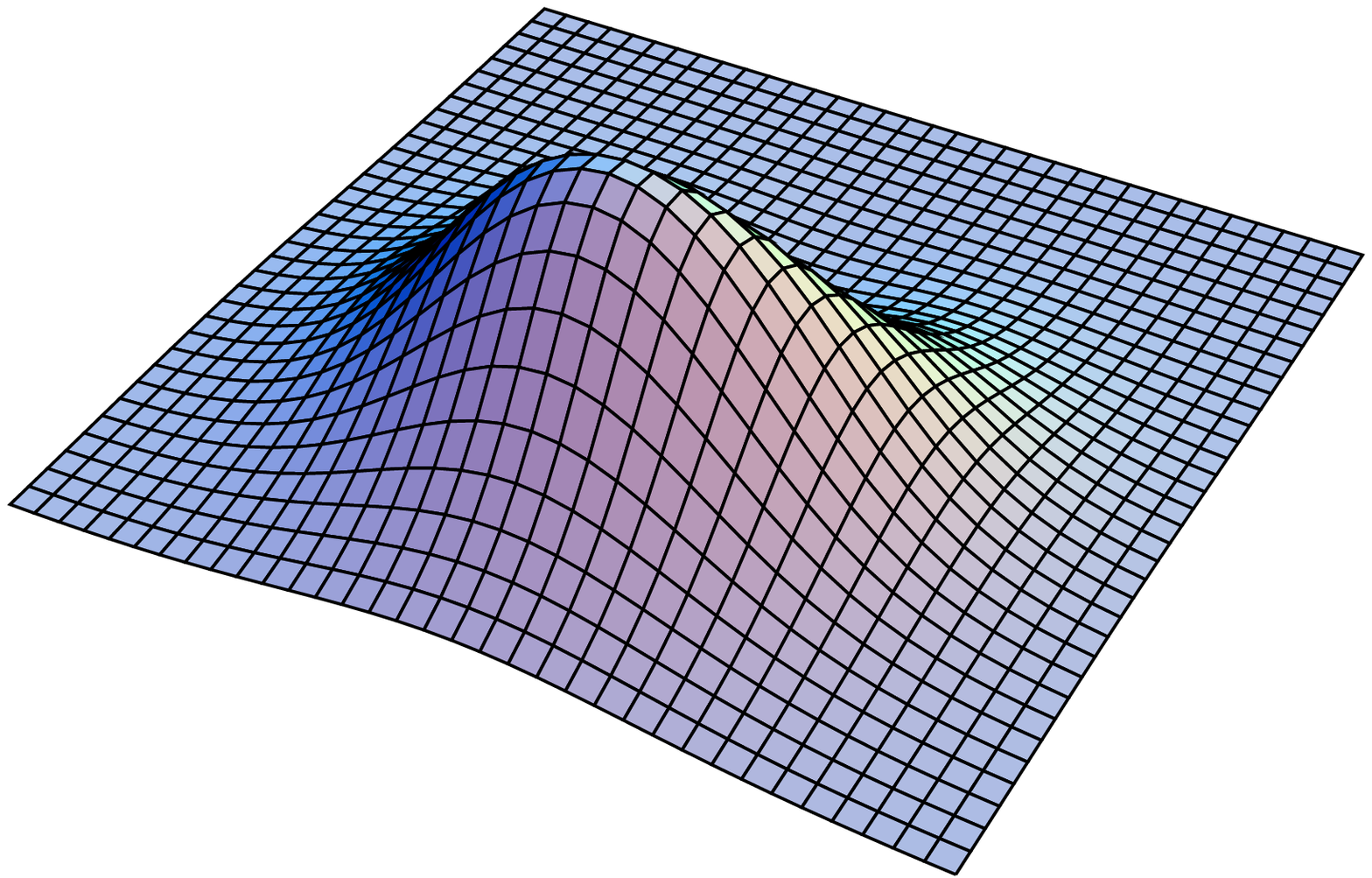}
\includegraphics{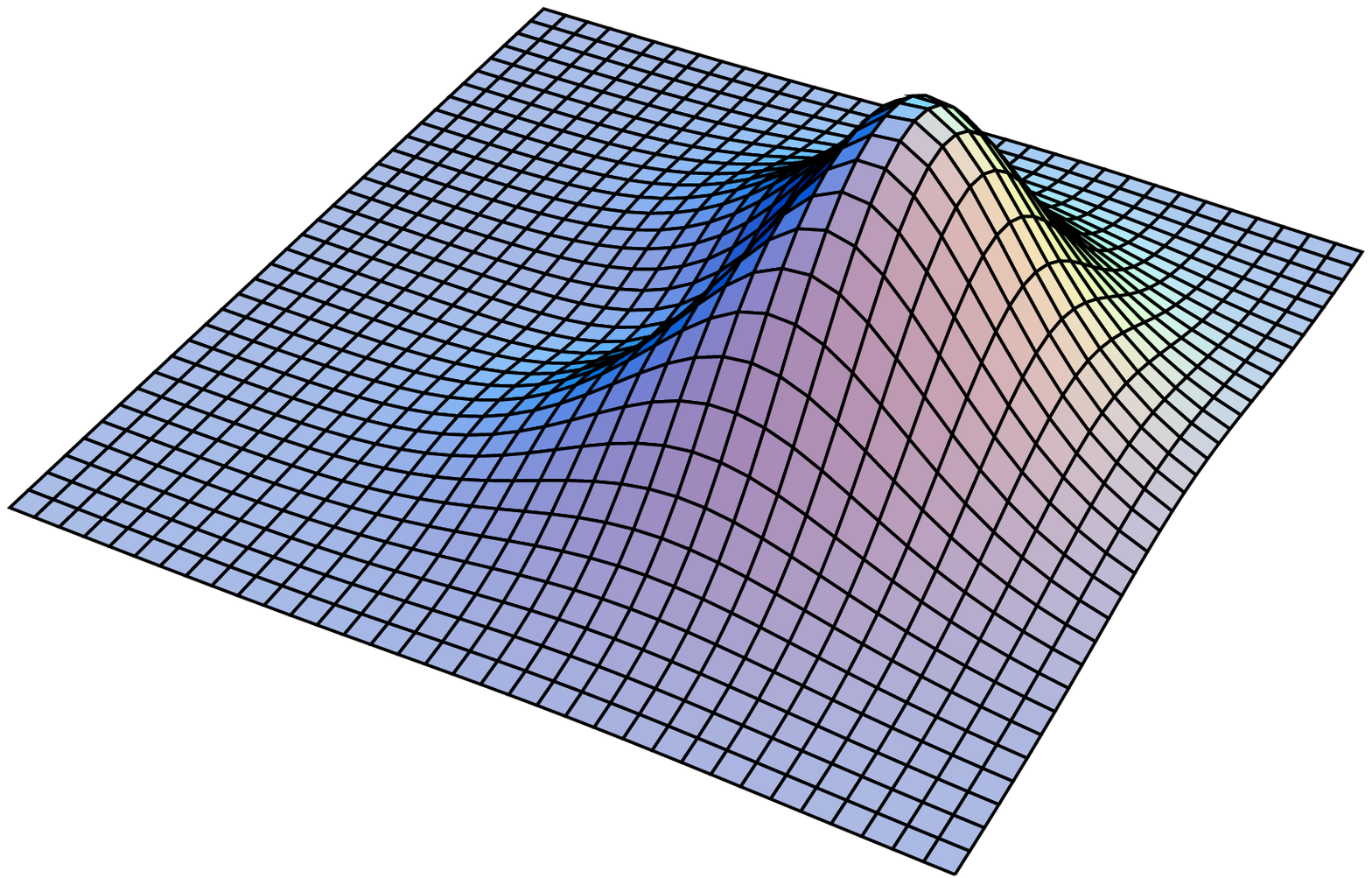}
\includegraphics{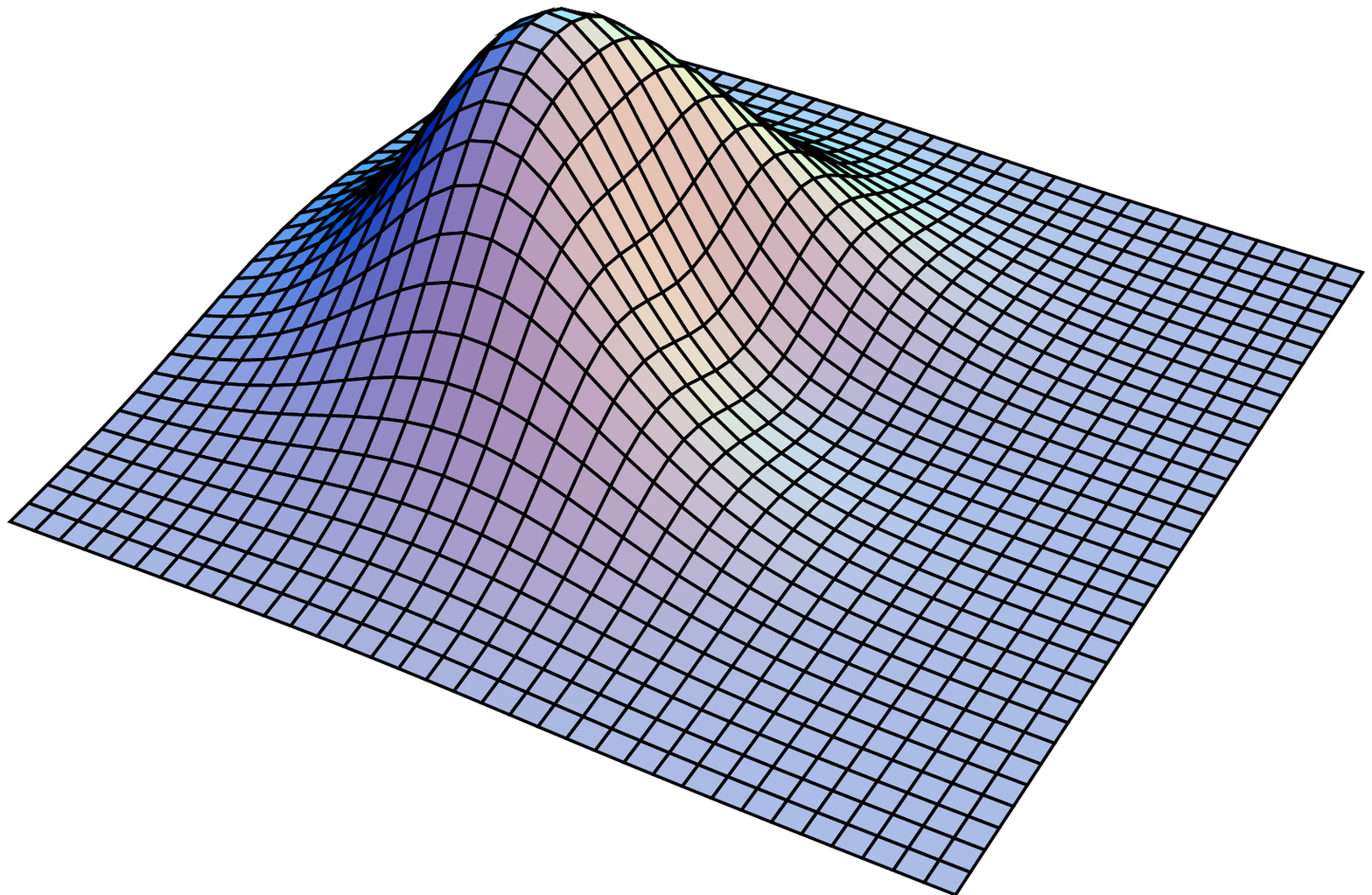}
\includegraphics{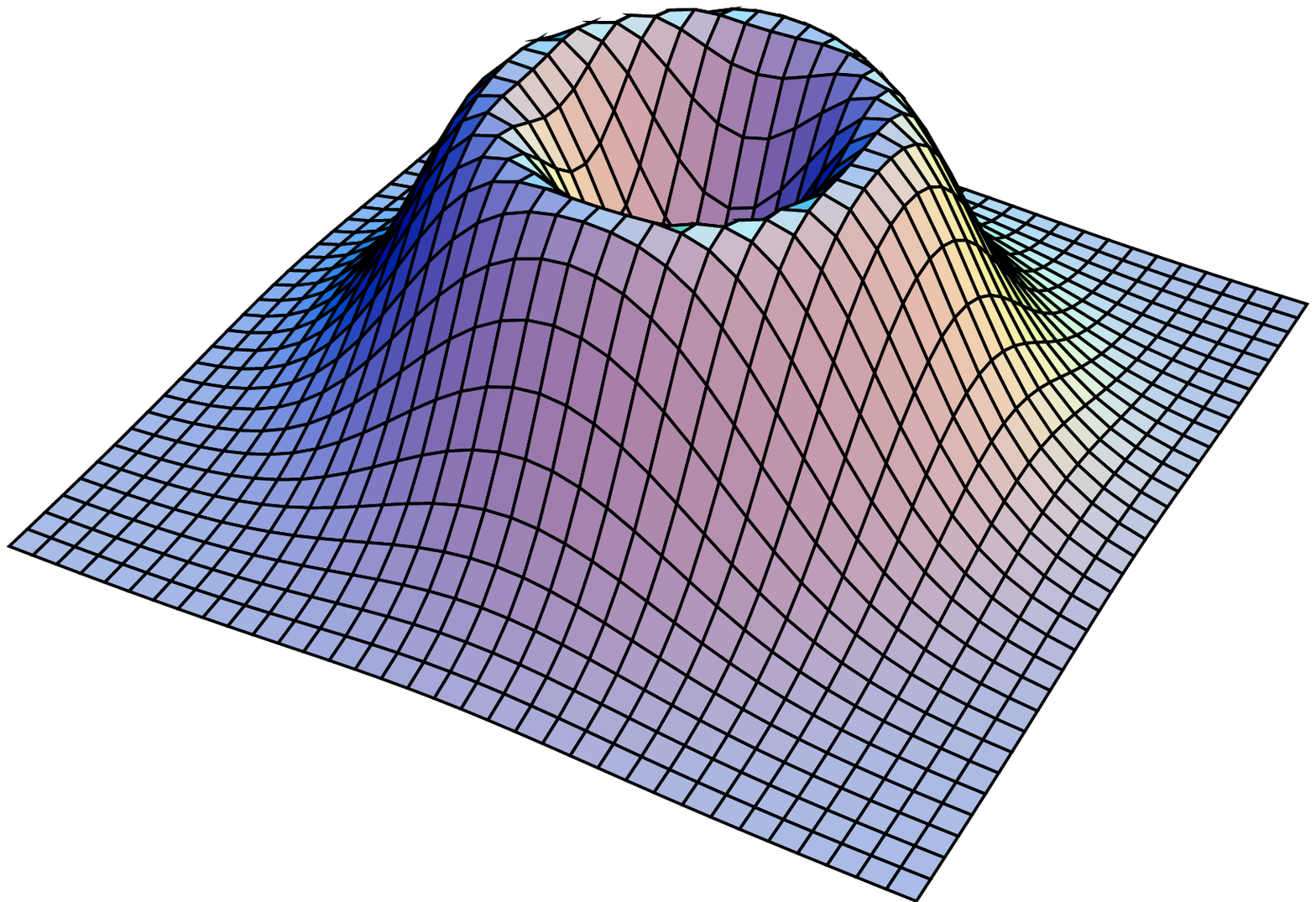}
\caption{In the middle is shown the action density in the plane of the
constituents at $t=0$ for an SU(2) charge 2 caloron with $\tr\,\pl=0$ in the 
crossed configuration of Fig.~\ref{fig:conloc}, hence with $\k=0.962$ and 
$D=3.894$. On a scale enhanced by a factor $16\pi^2$ are shown the densities 
for the two zero-modes, using either periodic (left) or anti-periodic (right) 
boundary conditions in the time direction.}\label{fig:close} 
\end{figure}

We will illustrate, using the exact formalism, how the action and zero-mode 
densities for these solutions behave. For the rectangular case we are mostly 
interested in the monopole limit, $d\to\infty$. For finite $d$, the density 
separates in two contributions, where each in the limit $d\to\infty$ is exactly
the density for a charge 2 monopole with the same values of $\k$ and $D$. Note 
that in the limit $d\to\infty$ the matching conditions for $\hat A_j(z)$ 
``decouple" the intervals, turning into pole conditions at $z=\mu_{1,2}$, as 
is appropriate for multi-monopoles~\cite{NahmM}. We do recover from this the 
known results for the charge 2 monopoles~\cite{MonRing}, like the doughnut 
structure for $\k=0$, corresponding to two superimposed monopoles. The interest 
in the monopole limit comes from the fact that the zero-mode densities for 
multi-monopoles had not been studied in detail before~\cite{CoGo}. In 
Fig.~\ref{fig:monlim} we give the densities for $\k=0.570$ and $D=6.915$, 
which is intermediate between the doughnut and well-separated monopole 
configurations. On the other hand, for $d$ small (compared to $\beta=1$) the 
configuration will look like two non-dissociated calorons, and in particular 
is no longer static. When $D$ remains much bigger than $d$, forcing $\k\to1$, 
these behave as two well-separated charge 1 instantons. Otherwise, when $D$ is 
comparable to $d$, one finds overlapping instantons~\cite{Overlap}.

An example for the crossed configuration with $\k=0.997$ and $D=8.753$ was 
already shown in Fig.~\ref{fig:fig1}. In this case $D$ is large enough for 
the like-charge constituents to be separated, as is particularly clear from 
the zero-modes, which are essentially no longer overlapping. But two nearest 
neighbour (oppositely) charged constituents still show appreciable overlap. 
The distance between these nearest neighbours is $\quart D\sqrt{2}/\pi=0.985$. 
As this is comparable to $\beta=1$ we would expect the configuration to depend 
on time. Indeed, at the maxima of the action density its value of $1.18
\times 16\pi^2$ at 
$t=0$ is reduced by almost 50\% at $t=0.5$. At the center of mass, 
where the action density is much lower, there is still a time dependence.
However, far from all cores the field becomes static\footnote{For any SU($n$) 
and topological charge $k$ the far field limit is static, {\em provided} all 
constituent monopoles have a non-vanishing mass}. Increasing $D$ further 
(which will push $\k$ closer to 1) the configuration quickly turns into 
well separated spherically symmetric static BPS monopoles. 

More interesting is to consider the case with smaller $D$, like $D=3.894$ 
and $\k=0.962$, for which the disc singularities were illustrated in 
Fig.~\ref{fig:conloc}. The corresponding densities are shown in 
Fig.~\ref{fig:close}. We see that the constituents are now so close 
that they form a doughnut, but we stress this is different from the 
{\em static} monopole doughnut, which has $\k=0$. Since the oppositely 
charged constituents now are as close as 0.438, which is considerably 
smaller than the time extent, the solution will have a strong time 
dependence. When $D$ is decreased even further, it will turn into a 
charge 2 instanton localized in both space and time.

\section{Higgs field asymptotics}\label{sec:Higgs}

In this section we make some comments on the far field limit of the gauge 
field. As we discussed before, the gauge field far removed from any core 
becomes abelian (as well as static). The abelian subgroup is the one that 
leaves the holonomy invariant, in the periodic gauge equivalent to leaving 
the constant asymptotic value of the adjoint Higgs field $A_0$ invariant. 
For definiteness, let us consider the case of SU(2) with $k=2$, $\beta=1$ and 
$\pl=\exp(2\pi i\vec\omega\cdot\vec\tau)$ (i.e. $\mu_2=-\mu_1=|\vec\omega|$). 
Up to exponential corrections we have 
\beq
A^\ff_0(\vec x)=2\pi i\vec\omega\cdot\vec\tau-\half i\hat\omega\cdot\vec\tau
\Phi(\vec x),
\eeq
where we can express $\Phi(\vec x)$ in terms of the far field limit of the 
Green's function at the impurities~\cite{BrvB}
\beq
\Phi(\vec x)=\pi^{-1}\left[1-\pi^{-1}\Tr\left(\hat f^{\,\ff}_x(\mu_2,\mu_2)
\hat S_2\right)\right]^{-1}\partial_i\Tr\left(\hat f^{\,\ff}_x(\mu_2,\mu_2)
\rho_2^i\right).
\eeq

Using the twistor description of magnetic monopoles Hurtubise~\cite{Hurt} was 
able to explicitly compute the asymptotic Higgs field for the SU(2) magnetic 
monopole long ago. The function he found for this algebraic tail amazingly 
agrees exactly with $\cV_m(\vec x)$, \refeq{Vdef}, which was introduced to 
describe the caloron zero-mode density ($m$ denotes the interval and hence 
the type of constituents to which the corresponding zero-modes would 
localize~\cite{Us}). As mentioned before, from our multi-caloron results one 
can recover the multi-monopole results by sending the constituent monopoles 
with the ``unwanted" magnetic charge to infinity, cmp. \refeq{Szm-alg}.

Although this tends to be cumbersome to show, $\Phi(\vec x)$ in the far field 
can be written as $\Phi_1(\vec x)-\Phi_2(\vec x)$, where $\Phi_m(\vec x)$ is 
the contribution coming from the type $m$ constituent monopoles and the 
difference in sign is due to the sign change in the magnetic charge. This 
is simply because the field is abelian in the far field and linear 
superposition preserves the self-duality. Hence, $\Phi_m(\vec x)=2\pi
\cV_m(\vec x)$, such that for the SU(2) caloron 
\beq
\Phi(\vec x)=2\pi\cV_1(\vec x)-2\pi\cV_2(\vec x).
\eeq
We checked that this result indeed holds for the solutions discussed in 
Sect.~\ref{sec:plots}, even when the two types of monopole structures are 
not well-separated. This relation trivially holds for the axially symmetric 
solutions that were introduced in Ref.~\cite{BrvB}, where $\Phi(\vec x)$ 
was explicitly shown to factorize in a sum of point charge contributions, 
compatible with what was found in Ref.~\cite{Us} for the zero-mode densities.
Therefore, in the far field limit (i.e. for the algebraic part) the 
singularity structure in the zero-mode density agrees {\em exactly} with 
the abelian charge distribution, as given by $\partial_i^2\Phi(\vec x)$. 
Such a relation is at the heart of using chiral fermion zero-modes as a 
filter to isolate the underlying topological lumps from rough lattice 
Monte Carlo configurations~\cite{Gatt}. 

\section{Discussions}

In this paper we have analyzed the higher charge caloron solutions and showed 
how to obtain exact results by suitably combining techniques developed in the 
context of the Nahm transformation and the ADHM formalism. The aim of these 
studies has been to establish that SU($n$) caloron solutions of charge $k$ 
can be described in terms of $kn$ monopole constituents, and that these can 
be viewed as independent constituents. A natural way to get an ensemble would 
be to consider approximate superpositions of $k$ charge 1 calorons, but this 
would lead to an unwanted memory effect, with constituents remembering from 
which caloron they originated~\cite{BrvB}. Our studies, within the context of 
self-dual configurations, have shown nevertheless that the constituents have 
an independent identity, with the only requirement that the net magnetic and 
electric charge of the configuration vanishes (each of the $n$ types of 
constituents should occur with the same number). A recent lattice 
study~\cite{Latt}, using the technique of over-improvement~\cite{OverImp},
fully confirms this picture. 

It is therefore reasonable to consider the constituents as the independent 
building blocks for constructing an ensemble of monopole constituents, 
something that was not questioned in Ref.~\cite{Diak}, but like for the 
instanton liquid~\cite{Shur} forms an essential assumption in a semi-classical 
study. Clearly the expectation is that semi-classical methods no longer work
in the confined regime, at least for the part of the parameter space that
corresponds to well-separated constituents, that is typically associated to 
instantons with a large scale parameter. It is not unlikely that the density 
of these constituents at low temperatures is so high that they form a coherent 
background and as such will no longer easily be recognized as lumps. With high 
quark densities leading to deconfinement, it may perhaps be that a high 
constituent monopole density will lead to confinement~\cite{Zak}, although 
for now we have to leave this as a speculation.

Instantons that overlap get deformed and depending on the relative gauge 
orientation tend to ``repel", i.e. inspecting the action density distribution 
they do not get closer than a certain distance~\cite{Overlap}. When 
deconstructing instantons in monopole constituents, interestingly only 
like-charge constituents will show this effect, manifesting itself through the 
extended core structure. For unlike charges, from the point of view of the 
abelian field, the configuration behaves as with linear superposition. 
If as a consequence of this all abelian charge is annihilated, it disappears 
through forming a small instanton (localized in space and time), which in the 
limit of zero size describes the boundary of the moduli-space. The interaction 
between constituents of opposite duality is more complicated~\cite{Berlin,DS}. 

In conclusion, calorons with non-trivial holonomy have revealed a rich
structure, incorporating traditional instanton physics, but allowing for 
gauge fields that inherit some essential features associated to a confining 
background not present in the traditional formulations. The fact that the 
underlying constituents are monopoles opens the way to describe the confining 
aspects of the theory in terms of these degrees of freedom. Much work 
remains to be done when it comes to understanding the dynamics, but 
we hope to have convinced the reader that a consistent picture is 
developing that holds considerable promise for the future.

\section*{Acknowledgments}

We thank Conor Houghton for pointing us to Hurtubise's paper, and Chris 
Ford for discussions on the analytical aspects. We are grateful to 
Christof Gattringer, Michael Ilgenfritz, Boris Martemyanov and Michael 
M\"uller-Preussker for stimulating discussions and correspondence on 
lattice implementations. The research of FB is supported by FOM. He is 
grateful to Chris Ford and Conor Houghton for their hospitality while 
visiting Trinity college in Dublin, and thanks them, as well as Werner 
Nahm and Samson Shatashvili, for their interest.

\section*{Appendix A}

In this appendix we derive the zero-mode limit for the action density,
which assumes that the distance of $\vec x$ and the constituents of type 
$m$ to all constituents of type $m'\neq m$ is large, but where $\vec x$ 
and the constituent locations of type $m$ may otherwise be arbitrary.

As in Ref.~\cite{Us} we take $z=\mu_m+0$ for computing $\cF_z$ and use
that we can write $\det(\Eins_{2k}-\cF_{\mu_m})=\det(\Eins_{2k}-LK)$, 
where $K\equiv F_{m-1}\Theta_{m-1}\cdots\Theta_1\hat g^\dagger(1)F_n\Theta_n
\cdots\Theta_{m+2}F_{m+1}$ and $L\equiv\Theta_{m+1}F_m\Theta_m$, with 
\beqa
&&\Theta_m\equiv\pmatrix{\ein_k&\ein_k\cr 2\pi R^+_m(\mu_m)&-2\pi R^-_m
(\mu_m)\cr}^{-1}\!\!T_m\pmatrix{\ein_k&\ein_k\cr 2\pi R^+_{m-1}(\mu_m)&-
2\pi R^-_{m-1}(\mu_m)\cr},\nonumber\\&&\nonumber\\&&F_m\equiv \pmatrix{
f^+_m(\mu_{m+1})f^+_m(\mu_m)^{-1}&0\cr 0&f^-_m(\mu_{m+1})f^-_m(\mu_m)^{-1}\cr}.
\eeqa
We note the $K$ has {\em no} remaining dependence on the constituent
locations of type $m$. Writing $LK\equiv\hat L\hat K+\tilde L\tilde K$, with
\beq
\hat K\equiv\pmatrix{K_{++}&K_{+-}\cr 0&\ein_k\cr},\quad
\tilde K\equiv\pmatrix{0\quad&0\quad\cr K_{-+}&K_{--}\cr},\quad
\hat L\equiv\pmatrix{L_{++}&0\cr L_{-+}&0\cr},\quad
\tilde L\equiv\pmatrix{0&L_{+-}\cr0&L_{--}\cr}.
\eeq
we find $\det(\Eins_{2k}-LK)=\det(\hat K)\det(\hat K^{-1}-\hat L-\tilde 
L\tilde K\hat K^{-1})$. 

We next use 
\beq
\hat K^{-1}=\pmatrix{K^{-1}_{++}&-K^{-1}_{++}K_{+-}\cr 0&\ein_k\cr},\quad
\tilde K\hat K^{-1}=\pmatrix{0&0\cr K_{-+}K^{-1}_{++}&(K^{-1})_{--}\cr}
\eeq
and note that in the zero-mode limit $K^{-1}_{++}$, $K^{-1}_{++}K_{+-}$,
$K_{-+}K^{-1}_{++}$ and $(K^{-1})_{--}$ are exponentially small (cmp. 
Ref.~\cite{Us}, App. A), such that
\beq
\det\left(ie^{-\pi ix_0}(\Eins_{2k}-LK)\right)=\det(e^{-2\pi ix_0}K_{++})
\det(L_{++}).
\eeq
With the definition of $L$ we now find
\beq
L_{++}=\quart R_{m+1}^{-1}(\mu_{m+1})\left(R_{m+1}^-(\mu_{m+1})+S_{m+1}
\right)\tilde U_m\left(R_{m-1}^+(\mu_m)+S_m\right),
\eeq
where
\beqa
\hskip-8mm&&\tilde U_m\!=\!\cZ_{m+1}^+f_m^+(\mu_{m+1})f_m^+(\mu_m)^{-1}R_m^{-1}
(\mu_m)\tilde\cZ_m^+-\cZ_{m+1}^-f_m^-(\mu_{m+1})f_m^-(\mu_m)^{-1}R_m^{-1}
(\mu_m)\tilde\cZ_m^-,\nonumber\\ \hskip-8mm&&\cZ^\pm_m\!=\!\Eins_k\pm(R_m^-(
\mu_m)+S_m)^{-1}R_{m-1}^\pm(\mu_m), ~\tilde\cZ^\pm_m\!=\!\Eins_k\pm R_m^\mp
(\mu_m)(R_{m-1}^+(\mu_m)+S_m)^{-1}\!\!.\label{eq:Szmlim}
\eeqa
and $\tilde U_m$ contains {\em all} contributions due to the constituent 
locations of type $m$, up to {\em exponential} corrections in the distance 
of these, {\em and} of $\vec x$, to the other constituents. Hence $\log\det
\left(ie^{-\pi ix_0}(\Eins_{2k}-LK)\right)$ splits into the sum of two 
contributions, $\log\det(\tilde U_m)$ and $\log\det\bigl[\quart(R_{m-1}^+
(\mu_m)+S_m)e^{2\pi ix_0}K_{++}R_{m+1}^{-1}(\mu_{m+1})(R_{m+1}^-(\mu_{m+1})
+S_{m+1})\bigr]$, where the last term only depends on the constituent 
locations of type $m'\neq m$ whose contribution will decay inversely 
proportional to the fourth power of their distance. Allowing for 
{\em algebraic} decay (or in the monopole limit, sending all constituents 
of type $m'\neq m$ to infinity) such that in addition $\cZ_{m+1}^\pm=
\tilde\cZ_m^\pm=\Eins_k$, one thus finds \refeq{Szm-alg}.

A simple way to derive the result for the far field limit in \refeq{Sff} is 
by noting that in this case {\em all} $f_m^-(\mu_{m+1})f_m^-(\mu_m)^{-1}$ are 
exponentially small and $F_m$ can be approximated by $\diag(F_m^{++},0)$, 
with $F_m^{++}=f_m^+(\mu_{m+1})f_m^+(\mu_m)^{-1}$. This therefore acts as 
a projection on the $++$ component and is thus seen to lead to $\det(ie^{-\pi 
ix_0}(\Eins_{2k}-LK))=\det(e^{-2\pi ix_0}\hat g^\dagger(1)F_n^{++}
\Theta_n^{++}\cdots F_1^{++}\Theta_1^{++})$. Using the fact that~\cite{Us} 
$\Theta_m^{++}=\half R_m^{-1}(\mu_m)\Sigma_m$ gives the required result.

\end{document}